\documentclass[usenatbib,usegraphicx,usedcolumn]{mn2e}
\usepackage{bm}
\usepackage{amssymb,amsmath}
\usepackage{graphicx}
\usepackage{natbib}
\usepackage{program}
\usepackage{color}
\usepackage{multirow}

\usepackage[breaklinks,colorlinks,citecolor=blue]{hyperref}

\title[Dense shocked regions in MHD turbulence]{Star formation from dense shocked regions in supersonic isothermal magneto-turbulence}

\author[P. Mocz \& B. Burkhart]{Philip Mocz$^{1}$\thanks{E-mail: pmocz@astro.princeton.edu (PM)}\thanks{Einstein Fellow} \& 
Blakesley Burkhart$^{2}$
\\
$^{1}$Department of Astrophysical Sciences, Princeton University, 4 Ivy Lane, Princeton, NJ, 08544, USA \\
$^{2}$Harvard-Smithsonian Center for Astrophysics, 60 Garden st. Cambridge, MA, 02138,  USA
}

\begin{document}

\date{submitted to MNRAS, May 2018}

\pagerange{\pageref{firstpage}--\pageref{lastpage}} \pubyear{2018}

\maketitle

\label{firstpage}

\begin{abstract}
Supersonic isothermal turbulence establishes a network of transient dense shocks that sweep up material and have a density profile described by balance between ram pressure of the background fluid versus the magnetic and gas pressure gradient behind the shock. These rare, densest regions of a turbulent environment can become Jeans unstable and collapse to form pre-stellar cores. Using numerical simulations of magneto-gravo-turbulence, we describe the structural properties of dense shocks, which are the seeds of gravitational collapse,  as a function of magnetic field strength. In the regime of a weak magnetic field, the collapse is isotropic. Strong magnetic field strengths lead to significant anisotropy in the shocked distribution and collapse occurs preferentially parallel to the field lines. Our work provides insight into analyzing the magnetic field topology and density structures of young protostellar collapse, which the theory presented here predicts are associated with large-scale strong shocks that persist for at least a free-fall time.
\end{abstract}

\section{Introduction}\label{sec:intro}

Filament formation in molecular clouds appears to be fundamentally linked to star formation as filaments are the sites of pre-stellar cores \citep{2010A&A...518L.102A,Arzoumanian2011A&A...529L...6A,Hacar2011A&A...533A..34H,Hennebelle2013A&A...556A.153H,Gomez2014ApJ...791..124G,Schleicher2018MNRAS.475..121S,Chira2018A&A...610A..62C,Veena2018ApJ...852...93V}.  Ultimately, the density structure in star forming regions is set by the properties of gas, which, to first order is a supersonic isothermal medium, threaded by magnetic fields. 
The understanding of star formation, under such conditions, is also key to the understanding of galaxy evolution \citep{Elmegreen2004,Mckee_Ostriker2007,krumreview2014}.

Indeed, it has long been known, since Larson's relations \citep{1981MNRAS.194..809L,Solomon87a,Heyer2004,Ballesteros-Paredes11a},
that cloud velocity dispersions measured from linewidths indicate supersonic bulk motions that show a general hierarchy as a function of length scale characteristic of a cascade of motions in turbulence \citep{Lazssrv09,burkhart2013,2013MNRAS.436.3247K}.
Furthermore, polarization and Zeeman observations indicate the presence of magnetic fields that thread the ISM on diffuse and dense scales \citep{Goodman1995ApJ...448..748G,2010ApJ...725..466C,2012ARA&A..50...29C,2016A&A...586A.138P}.
The large-scale diffuse ISM shows density-independent magnetic field strengths, which may suggest
these diffuse clouds are assembled by flow along the magnetic field lines and/or the presence of turbulence \citep{Troland08a,Crutcher2009,LEC12}. In denser regions, the magnetic field (as measured by Zeeman splitting) is
observed to increase with density, growing due to the flux-frozen condition of the field lines \citep{2012ARA&A..50...29C}.
The relative orientation between density gradients and magnetic fields also changes behavior as a function of density \citep{2016A&A...586A.138P,2017A&A...607A...2S}.
At low column densities, as observed by polarization measurements from the \textit{Planck} satellite, the
magnetic field orientations are either parallel or unoriented with respect to density gradients, but at high column densities they change to perpendicular alignment. This transition may be an indication of turbulence that is (sub-)Alfv\'enic, meaning that the magnetic field influences gas dynamics. Observations of velocity gradients, which correspond to the magnetic field direction, also confirm the influence of magnetic fields in shaping the density of both diffuse and molecular clouds \citep{Laz2018ApJ...853...96L}. Finally, observations of molecular gas shows the environment out of which stars form is nearly isothermal on scales above $0.01$~pc, due to a combination of collisional heating and radiative cooling processes \citep{2001Natur.409..159A}.

Thus, to first order, stars form out of a supersonic magnetized isothermal environment, in which turbulence acts to stir density perturbations. Turbulence and magnetic fields both create support against collapse from self-gravity, making star formation inefficient \citep{2004RvMP...76..125M,Krumholz2005,Mckee_Ostriker2007,Padoan11b,Hennebelle11b,federrath12,2012MNRAS.423.2016H,krumreview2014,2018arXiv180105428B}.
The origins of stirring of gas turbulence may come from various sources, including gravitational instability from gas accretion onto galaxies, or supernova-driven turbulence, which have different impacts on the global distribution of velocity distributions in a galaxy \citep{2011ApJ...738..101G,2011MNRAS.411...65B,2014MNRAS.438.1552F,2016MNRAS.458.1671K,Ibanez2017ApJ...850...62I,Krumholz2018MNRAS.tmp..831K}. Many processes beyond magnetic/turbulent support and density enhancements from supersonic motions may play important roles in star formation. These include
radiative feedback and stellar outflows from young stars, as well as competitive accretion and dynamical interactions between young stars \citep{2008ApJ...681..771M,2015ASSL..412...43K,2016ApJ...832...40K}. For example, simulations have shown that a massive star's radiation field and stellar winds are important mechanisms for regulating star formation \citep{goodman09,Offner09b,2016MNRAS.463.2553R}. Furthermore, on smaller scales  the isothermal condition breaks down as collapsing gas becomes optically thick to its own cooling radiation. Non-isothermal effects can affect the statistics of turbulence \citep{Federrath15a}. 

The focus of the present work is to understand the initial conditions and geometries out of which stars begin to form, and how large scale turbulent magnetic clouds set these properties. Thus we focus on the role of turbulence and magnetic fields, as the relative importance of their role in regulating star formation is still largely uncertain. The physical picture investigated here will also provide new ways to test the isothermal turbulent fragmentation model and place constraints on the Alfv\'enic Mach number from observations \citep{EL11,burkhart14,2018ApJ...853...96L}.

Using the statistical properties of a turbulent density field, there has been theory developed to predict the
initial mass function (IMF) of collapsing cores and star formation rates \citep{2002ApJ...576..870P,2012MNRAS.423.2016H,Padoan11b,Hennebelle11b,federrath12,Zamora2014ApJ...793...84Z,2018arXiv180105428B}. Diffuse turbulently stirred gas shows shows a log-normal probability distribution of densities (1 point statistic), whose width increases with the turbulent Mach number $\mathcal{M}$ \citep{Vazquez-Semadeni1994,Padoan1997,Krumholz2005,Hill2008,Burkhart2012}. Increased sonic Mach number produced density fluctuations which also flatten the power spectrum (2 point statistic) and produce non-Gaussian signals in the 3 point correlation function \citep{Kowal07a,Burkhart09a,burkhart10,2015ApJ...808...48B,Portillo2017}. At high densities, the gas is dense enough to be Jean's unstable and undergo gravitational collapse. This leads to a creation of a power-law tail in the distribution of dense gas, which is a behavior that is observed at high densities \citep{Kainulainen09a,Ballesteros-Paredes11a,Lombardi10a,Collins12a,federrath12,2012A&A...540L..11S,Kainulainen13b,Girichidis2014,MyersP2015,schneider2015MNRAS.453L..41S,Stutz2015A&A...577L...6S,Burkhart2015,Imara2016,padoan2017ApJ...840...48P,MyersP2017,Bialy2017ApJ...843...92B,Chen2017,Burkhart2017ApJ...834L...1B,2017ApJ...838...40M,2018arXiv180105428B}. 

The hypothesis that the IMF primarily originates as a consequence of supersonic
turbulence (ignoring stellar feedback) does produce reasonable predictions for star formation, where the dependence of the IMF turnover on physical parameters has been tested numerically \citep{2017arXiv170901078H}, and further testing of the model requires simulating and observing a large statistical sample with large dynamic range in space and time scales.
The model for the pre-stellar core IMF can be thought of as akin to an excursion-set model for dark matter halos for cosmic structure formation, for which fragmentation/merger trees can be constructed \citep{2012MNRAS.423.2016H}.
The understanding of the time evolution of the powerlaw part of the model, can also explain star formation 
that is spatially and temporally variable within a cloud, and depletion times of giant molecular clouds (GMCs) without the need to invoke feedback or extreme variations in the local star forming environments \citep{2018arXiv180105428B}.

The statistical picture of star formation from turbulent clouds relies on tracking a critical density for collapse \citep{Krumholz2005,Padoan11b,Hennebelle11b}.
In a similar spirit, critical density thresholds are used in cosmological simulations of galaxy formation and evolution to create star particles \citep{2018MNRAS.473.4077P}.
We address the critical density in a supersonic turbulent medium and its connection to star formation via the density PDF in a companion paper: Burkhart \& Mocz (2018). Here we are particularly interested to address the internal structure and time evolution of the seeds of collapsed regions. 
Detailed smaller scale simulations of star formation, which explore, for example, the effects of ambipolar diffusion 
\citep{2018MNRAS.473.3080M} or radiation \citep{2016MNRAS.463.2553R}, assume a collapsing core imposed by hand. 
These initial conditions typically fall into a class of critical Bonnor-Ebert spheres, may have laminar flow with angular momentum or a weakly turbulent velocity field imposed, and be threaded by uniform magnetic fields. 
However, a more accurate tracking of the collapse of the densest regions from general turbulent environment is important to enable a more complete picture for how the ISM gas transforms into stars, as pointed out recently by \cite{2018arXiv180105440R}. There is a missing gap of length scales in a complete theoretical understanding of the collapse process.

The work by \cite{2018arXiv180105440R} considers the structure and characteristic features of the densest regions in supersonic, isothermal hydrodynamic turbulence, which they propose would be the seeds of all gravitationally collapsing regions. \cite{2018arXiv180105440R} develop a simple model where the densest structures are a low volume-filling system of shocks which have a characteristic exponential profile in the post-shock region where the pressure gradient is in rough hydrostatic balance with the ram pressure of the gas being swept up by the shock (leading to an exponential atmosphere model). Isothermal jump conditions lead to density enhancements by a factor of the Mach number squared: $\mathcal{M}^2$. \cite{2018arXiv180105440R} identify such shocked regions with exponential profiles in numerical simulations to provide a conceptual picture of their distribution and time evolution. The work predicts, based on estimates for dissipation timescale vs free-fall timescale and fluctuation timescale of the potential, that some of these shocked regions would collapse under self-gravity, despite not containing substantial mass when considered as distinct regions, e.g. similar to the earlier proposals by \citep{Krumholz2005,Padoan11b,Burkhart2017ApJ...834L...1B}. These regions would be the seeds for star formation.
Often, as in cosmology, structure formation is described and predicted by looking at the Fourier power spectra of matter \citep{2003ApJS..148..175S}. But in the case of supersonic turbulence, where the medium is riddled with strong shocks, these discontinuities add equal power on all scales \citep{2015ApJ...808...48B}, and thus understanding structure formation is not completely amenable to analysis by spectral methods. Understanding the time evolution and statistics of these discontinuous shocks in physical space appears to be key in bridging the gap of collapse from turbulent clouds to pre-stellar cores.

Here we explore the the results discussed in \citep{2018arXiv180105440R} with self-gravitating turbulent simulations.  Furthermore, we explore how a magnetic field (self-consistently generated by turbulence) augments this picture. A strong magnetic field would be expected to lead to anisotropic structures, as flow along field-lines is easier than perpendicular to it. Furthermore, the if the field is strong enough, the flux-frozen condition of ideal magnetohydrodynamics (MHD) could prevent the collapse of a core completely along a certain direction. The work adds new insight into the structural evolution of dense collapsing regions, and how shocks remain associated with collapsing pre-stellar cores, which may be an important ingredient in understanding their further small-scale evolution.

The paper is organized as follows.
In Section~\ref{sec:sims} we describe the details of the suite of numerical simulations of supersonic, isothermal MHD turbulence with self-gravity.
Some 1D MHD simulations of shock formation are provided in Section~\ref{sec:1d}, useful for the interpretation of results.
Section~\ref{sec:3d} shows the analysis of our 3D simulations, including the identification and tracking of the properties of sites of pre-stellar core collapse.
A discussion of the results is offered in Section~\ref{sec:disc}, and we present our conclusions in Section~\ref{sec:conc}.

\begin{table}
\caption{Turbulent collapse simulation parameters}
\begin{center}
\begin{tabular}{ccccc}
\hline
\hline
sim. & $\beta_{\rm mean-field}$ & $\mathcal{M}$ & $\alpha_{\rm vir}$ & comment \\
\hline
1 & 25     & 10 & 1/2  & very weak field  \\
2 & 0.25   & 10 & 1/2  & weak field       \\
3 & 0.028  & 10 & 1/2  & moderate field   \\
4 & 0.0025 & 10 & 1/2  & strong field     \\
\hline
\hline
\end{tabular}
\end{center}
\label{tbl:sims}
\end{table}

\begin{figure}
\begin{center}
\begin{tabular}{ccc}
$\beta_{\rm mean-field}=25$ &  $\beta_{\rm mean-field}=0.25$ & \multirow{4}{*}{\includegraphics[width=0.07\textwidth]{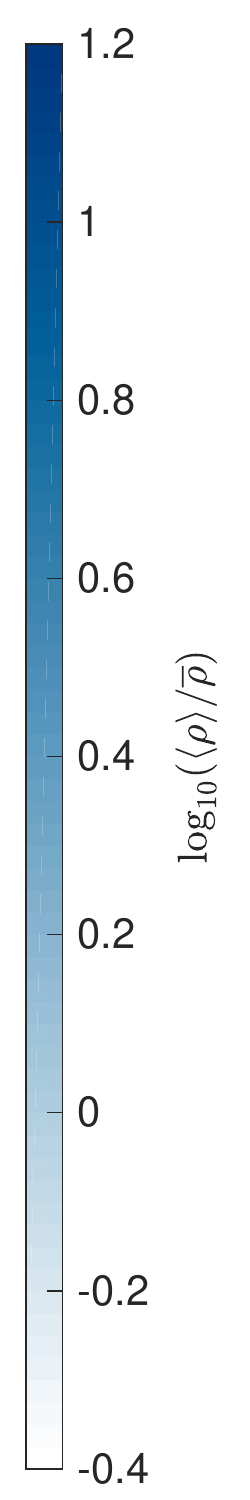}} \\
\includegraphics[width=0.175\textwidth]{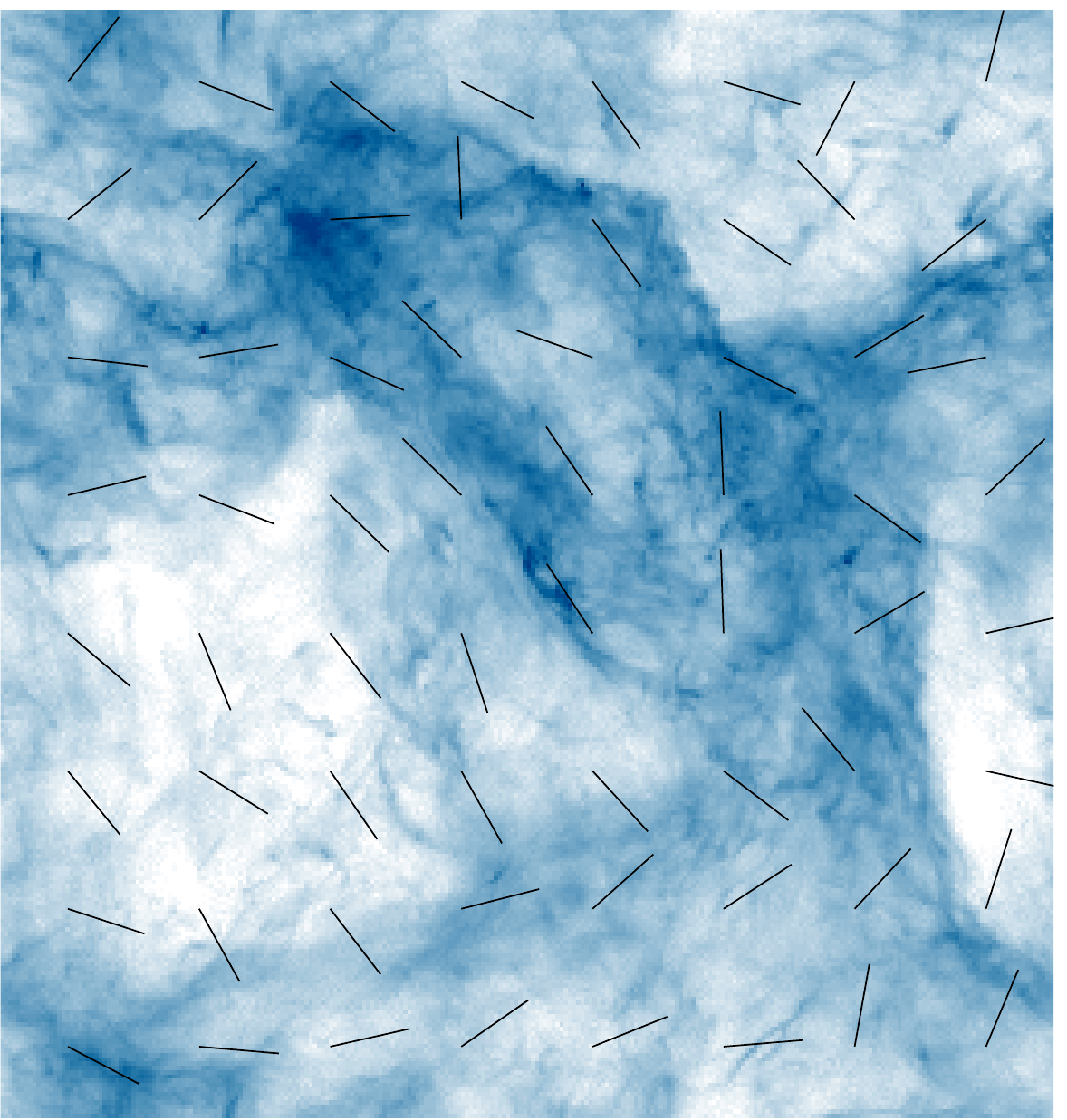} &
\includegraphics[width=0.175\textwidth]{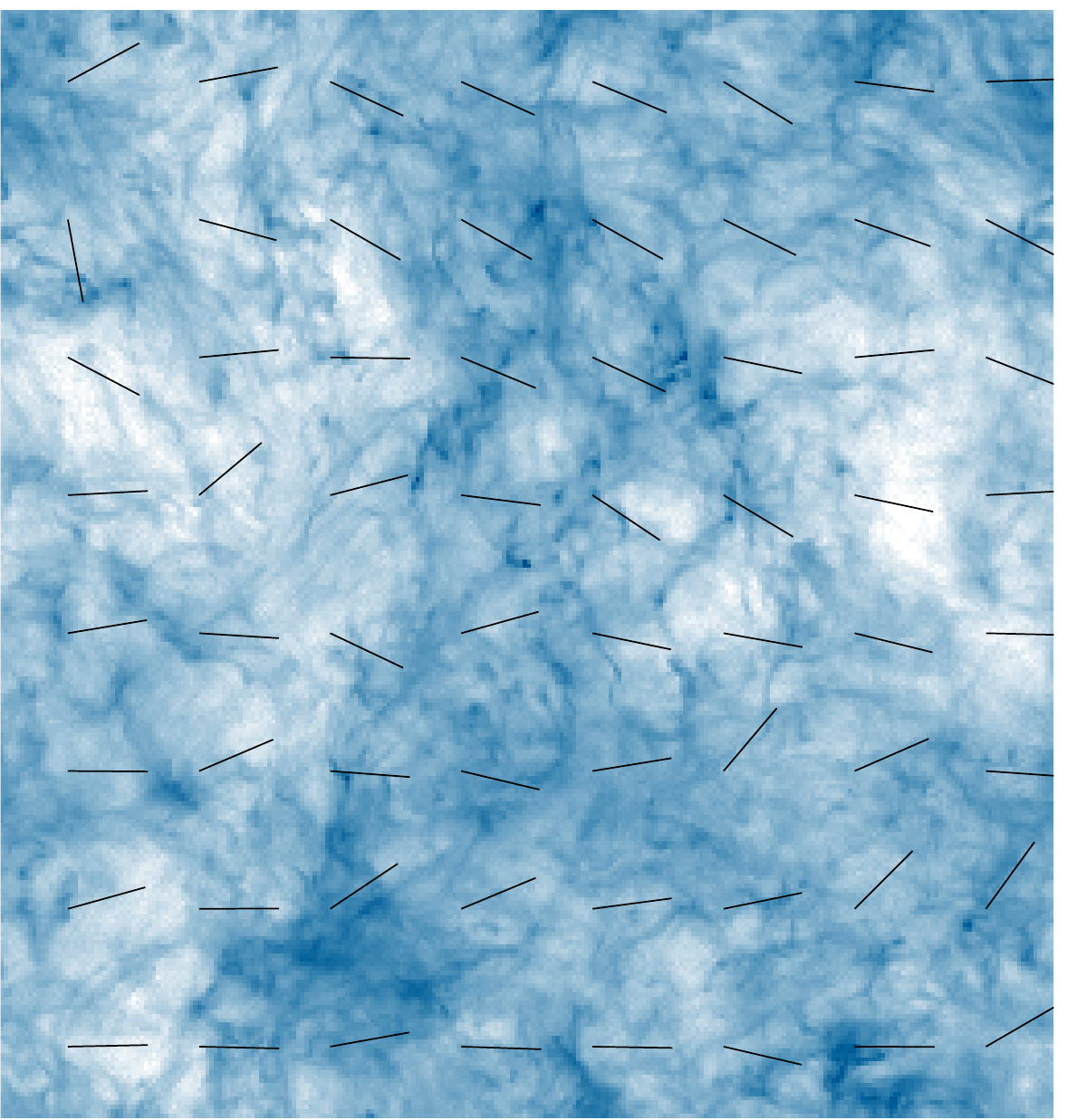} & \\
$\beta_{\rm mean-field}=0.028$ &  $\beta_{\rm mean-field}=0.0025$ & \\
\includegraphics[width=0.175\textwidth]{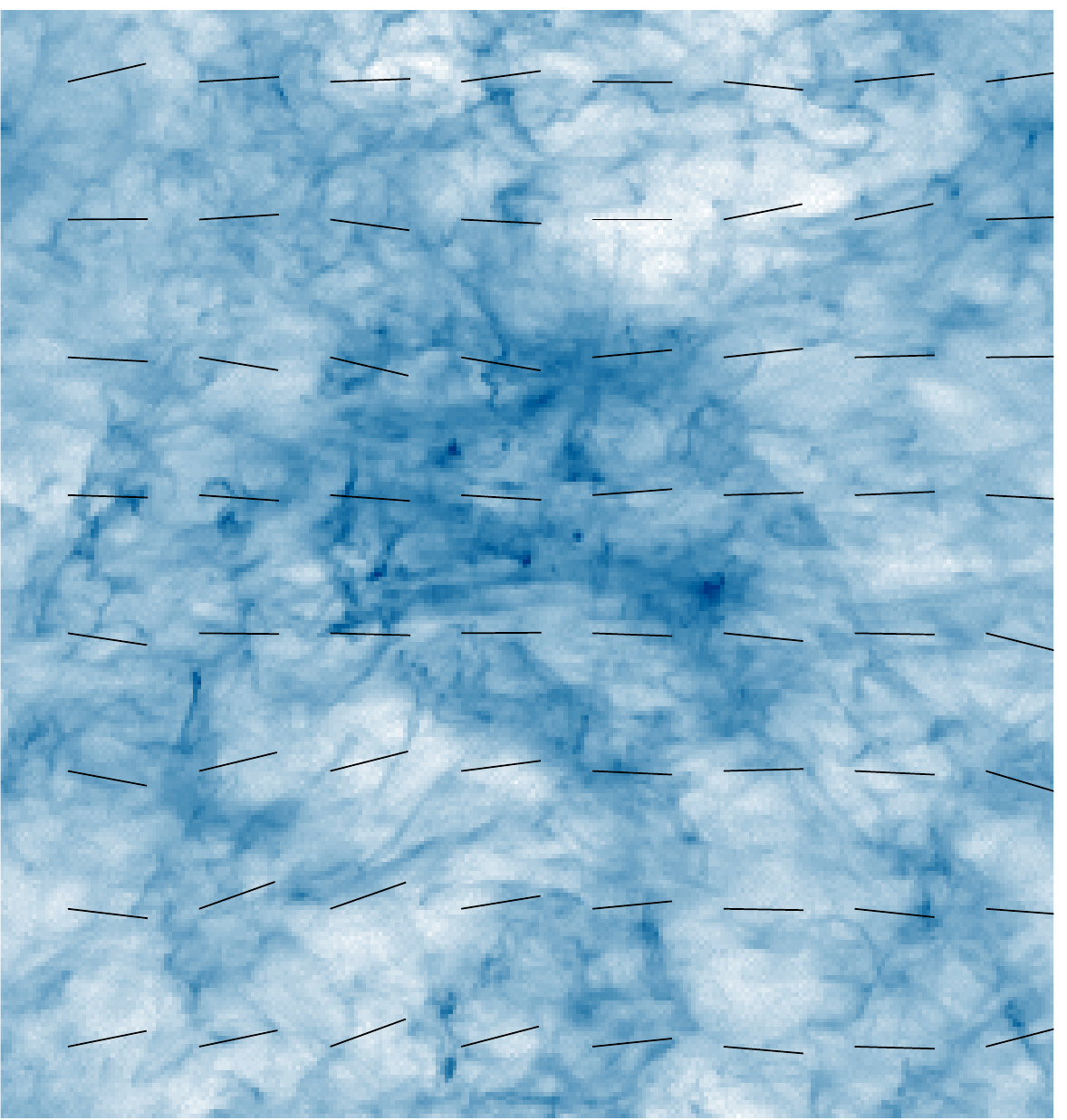}  &
\includegraphics[width=0.175\textwidth]{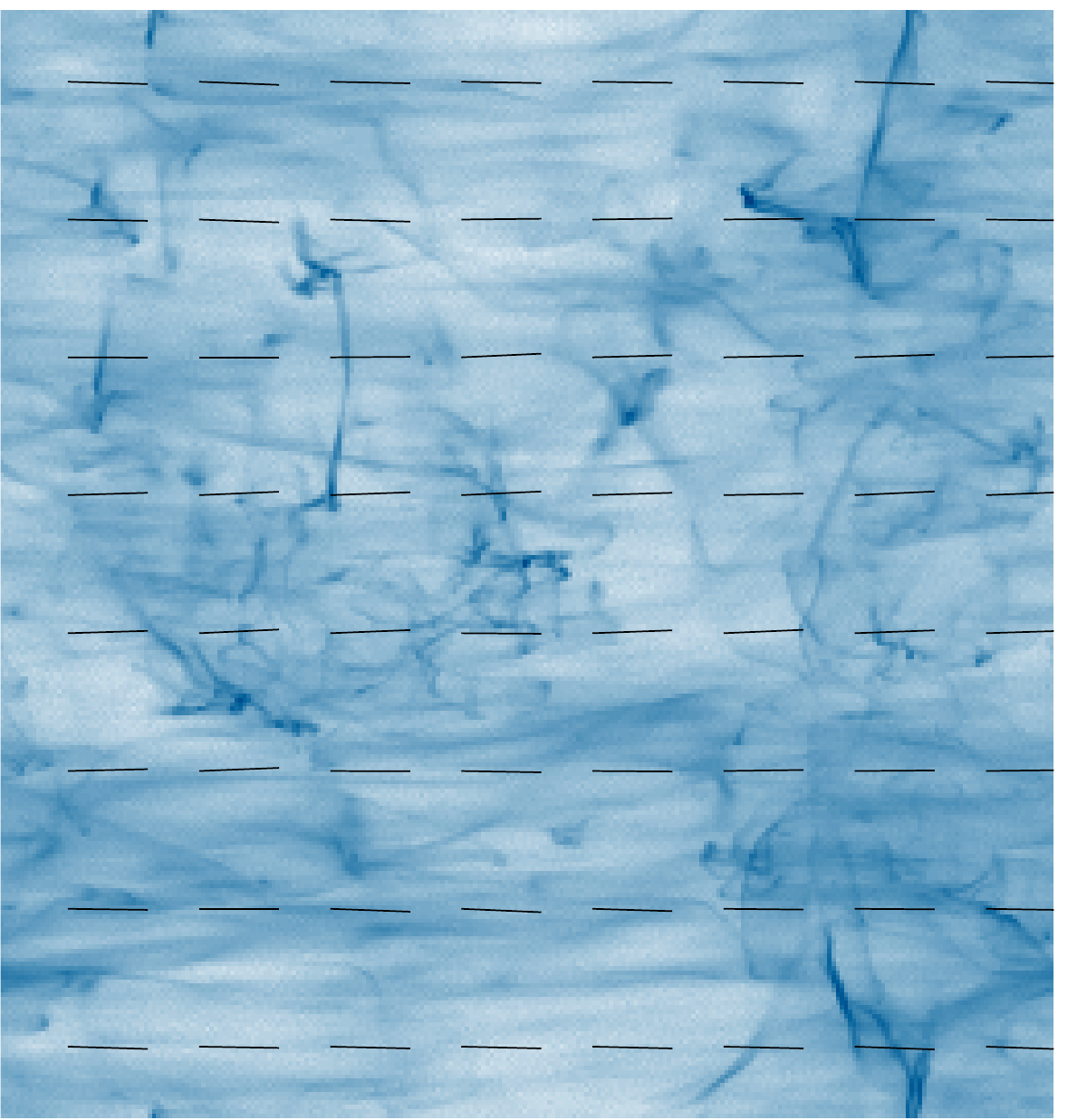} &
\end{tabular}
\end{center}
\caption{Projected densities and magnetic field vectors of the 3D turbulent self-gravitating simulations analyzed in this work. The strong field case is shown in the bottom right panel, with shocks organized perpendicular to the mean field clearly visible.}
\label{fig:sims}
\end{figure}

\begin{figure}
\begin{center}
\includegraphics[width=0.47\textwidth]{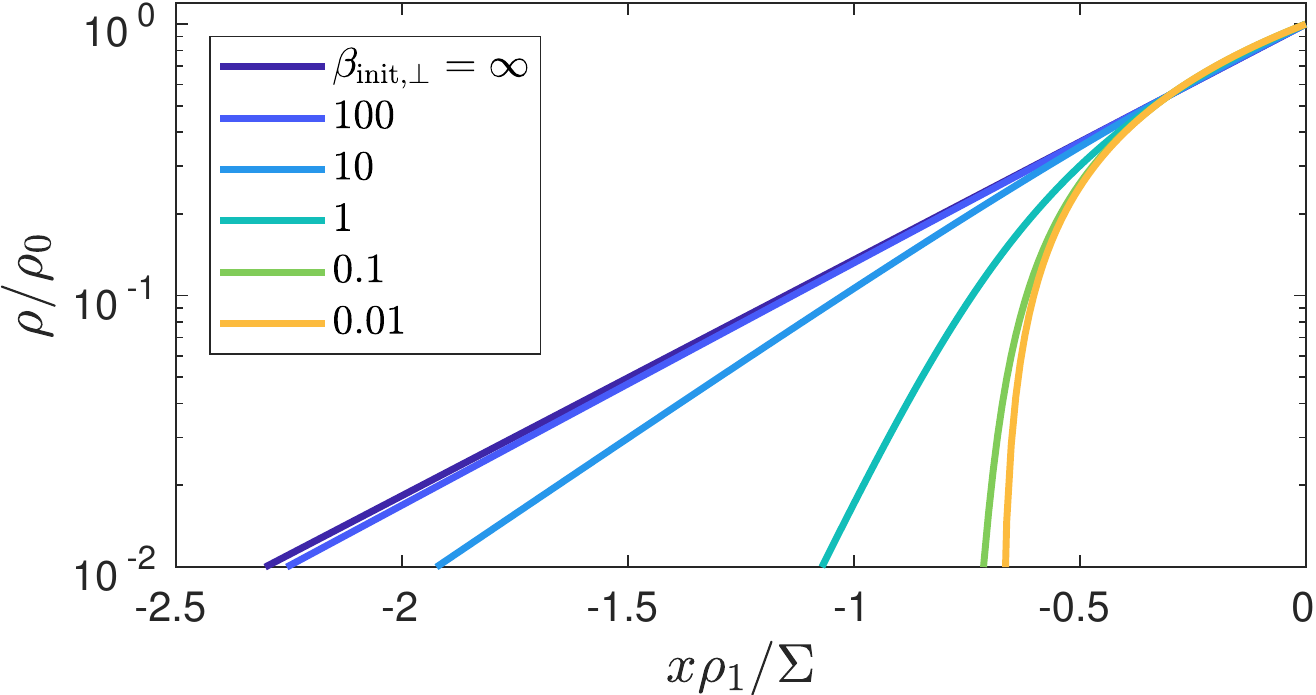}
\end{center}
\caption{Analytic perpendicular shock profile as a function of 
the initial perpendicular magnetic field strength. As the magnetic field goes to $0$, 
the solution reduces to an exponential atmosphere solution characterized by a scale-height $h$. Solution plotted here is for density jump $\rho_0/\rho_1=2$.}
\label{fig:1danalytic}
\end{figure}

\begin{figure}
\begin{center}
\includegraphics[width=0.47\textwidth]{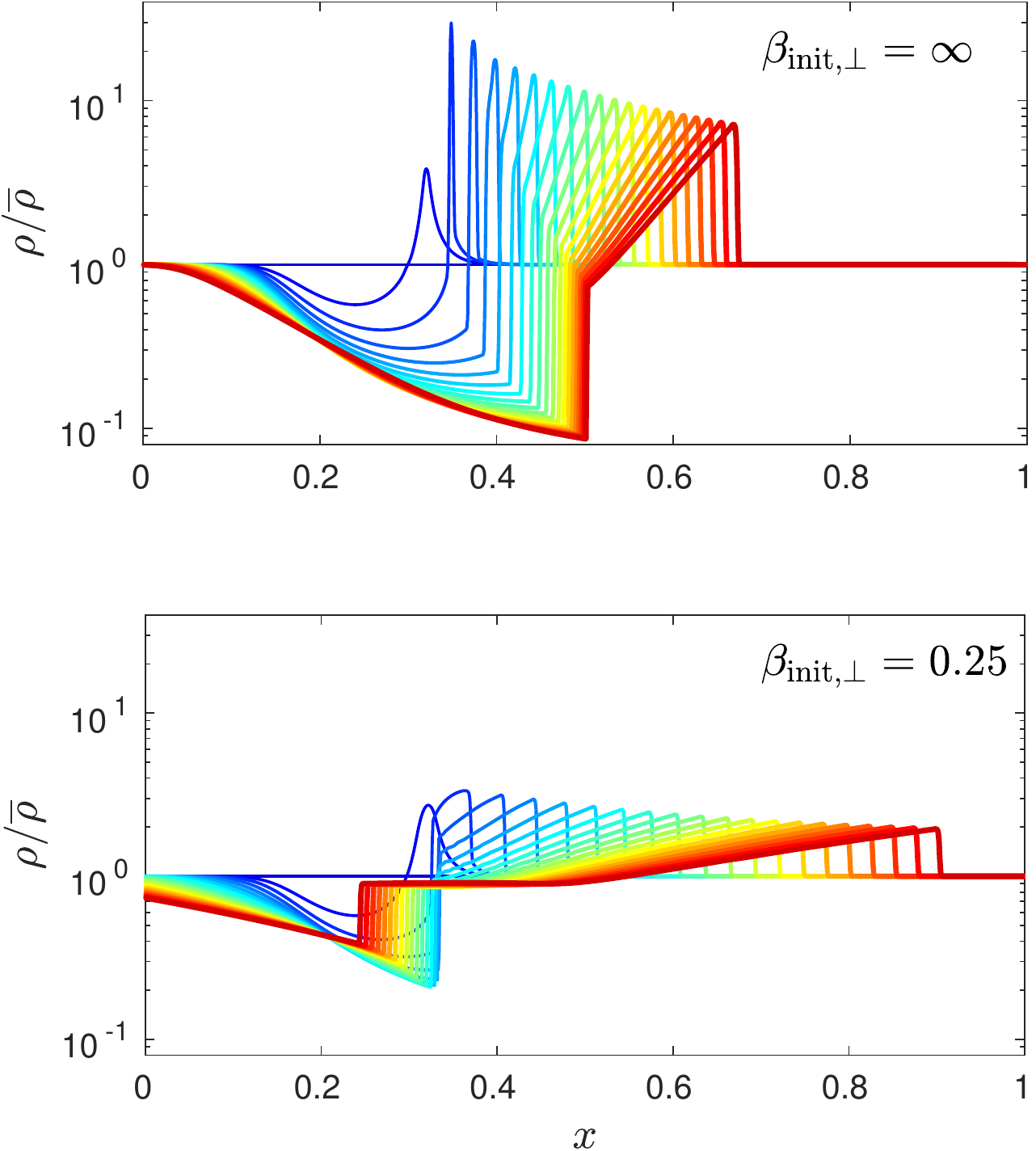}
\end{center}
\caption{Formation and expansion of a supersonic shock (20 snapshots of profiles from $t=0$ [blue] to $t=0.1 \tau_{\rm sound-crossing}$ [red] )
for the case of no magnetic field and a significant perpendicular magnetic field, in an idealized 1D simulation.
In the hydrodynamic case, the shock quickly forms an exponential profile, 
where the pressure gradient behind the shock balances the ram pressure of the swept up material. In the case of the parallel magnetized shock, 
the shock propagates at much faster speed, owing to the fast magnetosonic mode, 
and has a smaller density contrast but sweeps up more mass. The initial shock profile
is puffed up and not exponential due to the additional magnetic pressure support, but the solution 
evolves into an exponential atmosphere solution as peak density drops since the magnetic pressure drops with density.}
\label{fig:1dprofile}
\end{figure}

\begin{figure}
\begin{center}
\includegraphics[width=0.47\textwidth]{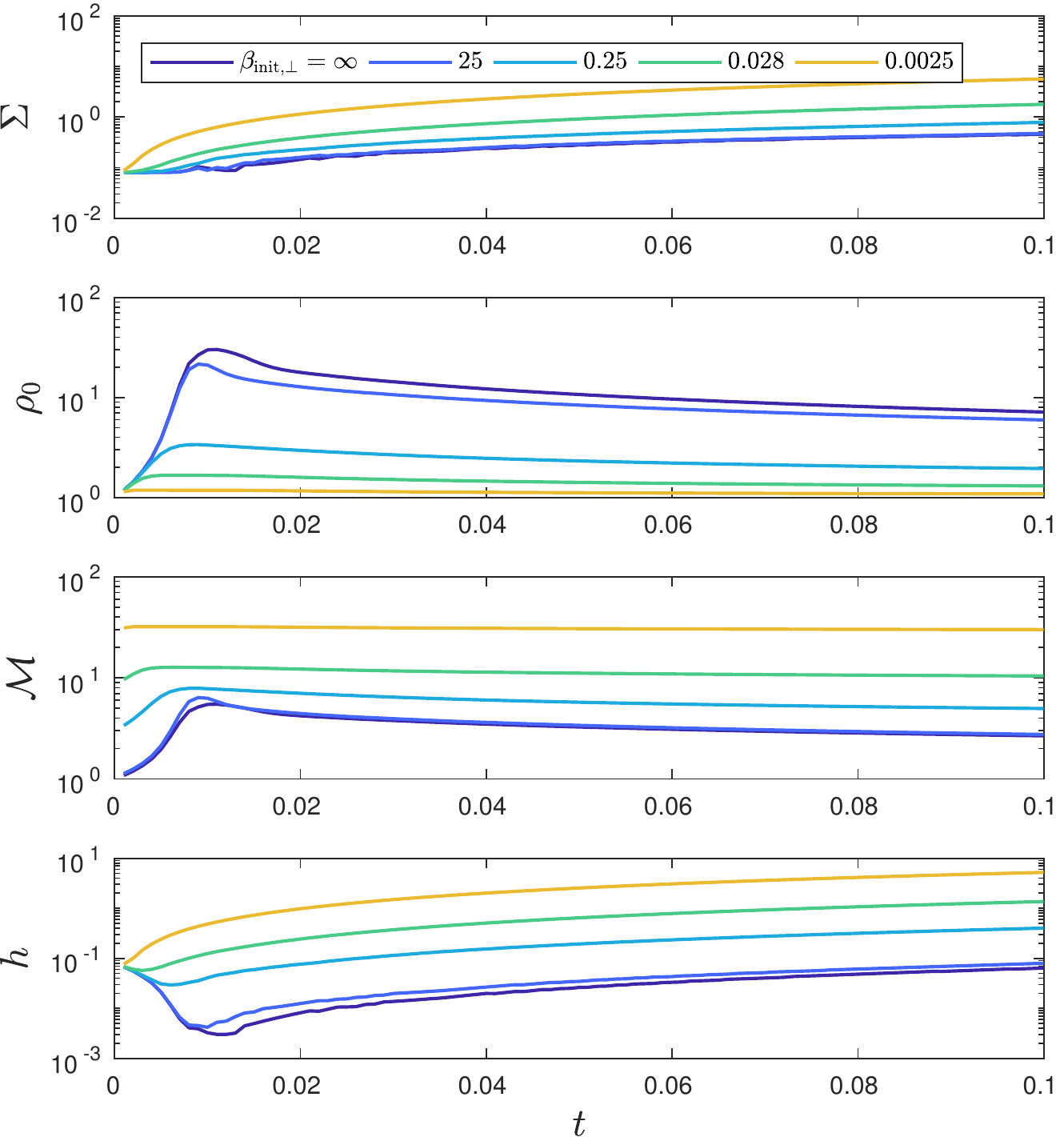}
\end{center}
\caption{The effect of the perpendicular magnetic field on a supersonic, isothermal shock. Being shown is the evolution of the mass per unit area of the shock, the peak density, the sonic Mach number, and the scale height evaluated as $h=\Sigma/\rho_0$.}
\label{fig:1dx4}
\end{figure}

\section{Turbulence simulations}
\label{sec:sims}

In this work we analyze a suite of supersonic, isothermal MHD turbulence simulations, 
with and without self-gravity, presented in \cite{2017ApJ...838...40M}.
The simulations were performed with the moving mesh {\sc Arepo} code \citep{2010MNRAS.401..791S}
which has a module for solving the MHD equations with constrained transport \citep{2016MNRAS.463..477M}
in order to maintain the divergence-free condition of the magnetic field.
The quasi-Lagrangian code has the advantage of resolving collapse and high density contrasts with high accuracy and minimal advection errors. Additionally, fluid parcels may be traced in a Lagrangian way in the simulation to learn about the origins of collapsed structure \citep{2013MNRAS.435.1426G}.

Table~\ref{tbl:sims} lists the simulation parameters studied. Projected densities and magnetic field vectors of the boxes are shown in Fig.~\ref{fig:sims}.

The turbulent simulations without self-gravity is characterized by just 2 parameters: 
the sonic Mach number $\mathcal{M}$ of turbulence, and the strength of the mean magnetic field (an invariant) 
characterized by the plasma beta $\beta_{\rm mean-field} = P_{\rm gas}/P_{B,\rm mean-field}$.
The simulations start out with uniform (dimensionless) density $\overline{\rho}=1$ in a box-size of $L=1$ and after a few eddy turnover times the system saturates to a quasi-steady state. 
The systems were driven solenoidally in velocity space on the largest spatial scales of the box with an Ornstein-Uhlenbeck process \citep{2010A&A...512A..81F,2012MNRAS.423.2558B,2015MNRAS.450.4035F}, which is a Gaussian and Markovian stochastic process.

The simulations include four isothermal turbulence simulations, 
representing part of a giant molecular cloud, run with different initial mean-field strengths, $B_0$. 
The sonic Mach number is set to $\mathcal{M} = v_{\rm rms}/c_{\rm s} = 10$, where $v_{\rm rms}$ is the root-mean-square velocity in the box and $c_{\rm s}$ is the sound speed, which may be typical of such star forming environments. The initial mean-field strengths have $\beta_{\rm mean-field} = 25,0.25,0.028, 0.0025$, 
ranging from weak to strong magnetic fields.

The simulation suite also includes four simulations in which self-gravity was switched on in a turbulent medium, after a quasi-steady turbulent state was reached. Gas was allowed to collapse for a time on order of the free-fall time.  The strength of gravity ($G$) in these dimensionless simulations sets a third scale to the problem, the Virial parameter of the cloud: $\alpha_{\rm vir} = 5 v_{\rm rms}^2 (L/2) /(3G M)$, where $M$ is the total mass in the box. These simulations have $\alpha_{\rm vir}=1/2$, representing a scale where self-gravity starts becoming important. In dimensionless units, the strength of gravity is $G=5\mathcal{M}^2/(6\alpha)$.

In this paper, we present results in terms of dimensionless quantities: e.g. overdensities $\rho/\overline{\rho}$, Mach numbers $\mathcal{M}$, fraction of box-size $x/L$. The simulations were performed with units of $\overline{\rho}=1$, $c_{\rm s}= 1$. We note though that the physical parameters of the simulations (assuming a mass per hydrogen of $1.4~{\rm amu}$) can be scaled as:
\begin{equation}
\begin{array}{l}
L = 5.2 \left(\frac{c_{\rm s}}{0.2~{\rm km}~{\rm s}^{-1}}\right)
\left(\frac{n_H}{1000~{\rm cm}^{-3}}\right)^{-1/2}
\left(\frac{\mathcal{M}}{10}\right)
 ~{\rm pc} \\
B_0 = 1.2,12,36,120 \left(\frac{c_{\rm s}}{0.2~{\rm km}~{\rm s}^{-1}}\right)
\left(\frac{n_H}{1000~{\rm cm}^{-3}}\right)^{1/2}
 ~{\rm \mu G} \\
M = 4860 \left(\frac{c_{\rm s}}{0.2~{\rm km}~{\rm s}^{-1}}\right)^3
\left(\frac{n_H}{1000~{\rm cm}^{-3}}\right)^{-1/2} 
\left(\frac{\mathcal{M}}{10}\right)^3 
 M_\odot \\
 \end{array}
\end{equation}
We scale the simulations to physical units using sound speed $c_{\rm s}=0.2~{\rm km}~{\rm s}^{-1}$ and hydrogen density $n_{\rm H} = 1000~{\rm cm}^{-3}$. The choice of physical scaling also makes the cloud 
fall on the observed line width-size scaling relation for molecular clouds in our Galaxy:
$\sigma_{\rm nt} = \sigma_{\rm pc} R_{\rm pc}^{1/2}$, with $\sigma_{\rm pc}\simeq 0.72~{\rm km}~{\rm s}^{-1}$ \citep{2007ARA&A..45..565M}. Here $R_{\rm pc}=(L/2)/(1~{\rm pc})$ and $\sigma_{\rm nt} = \mathcal{M}c_{\rm s}/\sqrt{3}$ (as in \citealt{2010ApJ...720.1612M}).

The system is characterized by timescales, 
where we also report their values in dimensionless units.
The sound crossing time:
\begin{equation}
\tau_{\rm sound-crossing} = \frac{L}{c_{\rm s}} = 1
\end{equation}
The eddy turnover time (also the Mach crossing time, and the timescale over which the gravitational potential changes):
\begin{equation}
\tau_{\rm eddy-turnover} \simeq \frac{L}{\mathcal{M}c_{\rm s}} = 0.1
\end{equation}
The free-fall time:
\begin{equation}
\tau_{\rm free-fall} \simeq \sqrt{\frac{3\pi}{32 G\overline{\rho}}} = \sqrt{\frac{\pi}{320\alpha}}\frac{1}{\mathcal{M}} = 0.014
\end{equation}

The simulations with self-gravity were run for a fraction of a free-fall time to follow the evolution of the structure of the first cores that form, covering a density contrast of 10 orders of magnitude, and following the entire stage of the isothermal collapse process. Beyond these scales, the
opacity increases to the point it would trap heat and hence the collapse would continue adiabatically, and the scalable nature of the process is broken.
Importantly, we have simulated the system with a resolution such that the Truelove criterion for resolving fragmentation due to the Jeans instability is met \cite{1997ApJ...489L.179T} at all times, without need for refinement beyond the Lagrangian spatial adaptability provided by the code down to the target density achieved. This means we capture all Lagrangian velocity perturbations in the collapse, which are, to zeroth order, frozen in to the fluid in a free-fall process. 
As we find in \cite{2017ApJ...838...40M}, the cores have significant turbulent kinetic energy as they collapse (e.g. see Figure 5 of that work).

\section{1D isothermal MHD shocks}
\label{sec:1d}

We describe and review some properties of simple 1D isothermal MHD shocks, which 
provides some basic intuition and proves useful later in our explanations of the structure of the densest regions of supersonic isothermal turbulence. These 1D structures are also found to approximately describe the starting conditions out of which pre-stellar cores begin to collapse due to self-gravity.

We consider a simple 1D setup 
where a uniform fluid of density $\overline{\rho}=1$ 
is given an initial supersonic kick in its velocity structure
according to a Gaussian pulse with peak velocity $v_{\rm peak}/c_{\rm s} = 10$:
\begin{equation}
v(x)/c_{\rm s} = 10 \times \exp\left(-\frac{(x-0.25)^2}{2\times 0.04^2}\right).
\end{equation}
This simple setup approximates focusing on the consequence of a single mode of large-scale stirring in 3D supersonic driven turbulence.
The supersonic motion of the gas parcels will lead to fluid characteristics to cross and the formation of a discontinuity -- a shock -- in the the fluid variables. The ``pre-shock'' region refers to unperturbed area ahead of the shock and the ``post-shock'' region refers to the processed fluid behind the shock.
In our setup, the ``post-shock'' is indexed by the subscript $0$ and the ``pre-shock'' is indexed by the subscript $1$, 
following the convention used in \cite{2018arXiv180105440R}.

We consider parallel and perpendicular shocks, for simplicity. That is, the setup supposed here is a 1D shock tube initially threaded by a uniform magnetic field either parallel or perpendicular to the direction of motion of the shock.
For a discussion on intermediate oblique shocks, see \cite{2014ApJ...785...69C}.
The magnetic field strength is characterized by the plasma-beta parameter $\beta = P_{\rm gas}/P_B$
(ratio of gas to magnetic pressure)
which has initial value $\beta_{{\rm init},\{\perp,\parallel\}} = \frac{\overline{\rho} c_{\rm s}^2}{B_{\rm init}^2/(8\pi)}$
where the subscript $\perp$ or $\parallel$ indicates whether the magnetic field is in the perpendicular or parallel direction to the shock velocity.
We consider pure hydrodynamic shocks as well as shocks with weak through strong magnetic fields: $\beta_{\rm init}=\infty$, $25$, $0.25$, $0.028$, $0.0025$.

In the case of no magnetic field, sitting in the frame of the shock, a shock will develop with some Mach number $\mathcal{M}\equiv v_{\rm 1}/c_{\rm s}$ which leads to an $\mathcal{M}^2$ jump in the density contrast in the post- and pre-shock regions: $\rho_0/\rho_1 = \mathcal{M}^2$, where $\rho_0$ is the post-shock density and $\rho_1=\overline{\rho}$ is the pre-shock density.

In the case where the magnetic field is parallel to the shock
velocity and initially constant, the magnetic field stays constant by the flux-frozen condition and it drops out of the shock jump conditions thus the solution reduces to the case of a pure hydrodynamic shock.

In the case of the perpendicular shock, the shock that forms corresponds to the fast magnetosonic wave
and has velocity that exceeds the fast magnetosonic speed $v_{\rm f} = \sqrt{v_{\rm A,1}^2 + c_{\rm s}^2}$
where $v_{\rm A,1} = B_1/\sqrt{4\pi}$ is the Alfv\'en wave speed in the pre-shock region. By the flux-frozen condition, 
the magnetic field stays perpendicular and its strength is proportional to the fluid density.
Since in an isothermal fluid the gas pressure scales as $\rho$ and the magnetic pressure scales here as $B^2\propto \rho^2$, the relative importance of the magnetic field becomes larger at higher overdensities in the fluid. 
The shock jump condition is given by:
\begin{equation}
r\equiv \frac{\rho_0}{\rho_1} = \frac{ -(1+\beta_1) + \sqrt{(1+\beta_1)^2+4\beta_1 \mathcal{M}^2} }{2}.
\end{equation}
In the limit of weak field the hydrodynamical limit is recovered $\lim_{\beta_1\to\infty} r = \mathcal{M}^2$. In general having a stronger perpendicular magnetic reduces the density contrast.
The shock jump condition may also be solved to obtain the Mach number of the shock:
\begin{equation}
\mathcal{M} = \frac{v_1}{c} = \sqrt{\frac{\rho_0^2+\rho_0\rho_1(1+\beta_1)}{\beta_1 \rho_1^2}}
\end{equation}

The shocked region profile can be described by assuming an 
equilibrium between the ram pressure $\rho_1 v_1^2$ exerted by the oncoming material 
and the pressure gradient (magnetic and gas) in the shocked region, 
similar to the pure hydrodynamic analysis in \cite{2018arXiv180105440R}.
In the case of a parallel shock the magnetic field plays no role in the shock profile 
so we consider a perpendicular shock with an initial magnetic field strength 
given by $\beta_{{\rm init},\perp} = \beta_1$ and the magnetic field remains proportional to density: $B = \sqrt{8\pi c_{\rm s}^2/(\beta_1 \rho_1)}\rho$. Define $\Sigma=\int \rho \,dx$
the mass per unit area of the shocked region along the $x$-direction of travel.
Then, equating post-shock pressure gradient with the ram pressure gives:
\begin{equation}
\nabla (P_{\rm gas} + P_{B}) = -\rho\frac{\rho_1v_1^2}{\Sigma}
\label{eqn:pbal}
\end{equation}
or, equivalently,
\begin{equation}
\frac{d}{dx}\left(  \rho c_{\rm s}^2 + \frac{\rho^2c_{\rm s}^2}{\beta_1 \rho_1} \right) = -\rho\frac{\rho_1v_1^2}{\Sigma}
\end{equation}
which has solution:
\begin{equation}
\begin{split}
\rho(x) &= \\
&= \frac{\beta_1\rho_1}{2}  \times 
W\left[ 
\frac{2\rho_0}{\beta_1\rho_1} 
\exp\left(  
\frac{2\rho_0}{\beta_1\rho_1} 
- x \frac{\rho_1\mathcal{M}^2}{\Sigma} 
\right)
\right]
\\
&= \frac{\beta_1\rho_1}{2}  \times\\
& \,\,\,\,\,\,W\left[ 
\frac{2\rho_0}{\beta_1\rho_1} 
\exp\left(  
\frac{2\rho_0}{\beta_1\rho_1} 
- x \frac{\rho_0^2 + \rho_0\rho_1(1+\beta_1)}{\beta_1\rho_1 \Sigma} 
\right)
\right]
\end{split}
\end{equation}
where $W(z)$ is the Lambert-$W$ function (inverse function of $f(z)=z\exp(z)$).
In the limit $\beta_1\to\infty$, the solution reduces to an exponential atmosphere
$\rho(x) = \rho_0 \exp(-x/h)$ with scale height:
\begin{equation}
h = \frac{\Sigma}{\rho_0}.
\end{equation}
Fig.~\ref{fig:1danalytic} shows the analytic shock profile as a function 
of the initial perpendicular magnetic field strength.
In the limit of strong magnetic field strength, the solution reduces to:
\begin{equation}
\rho(x) = \rho_0\left( 1 - x \frac{\rho_0 + \rho_1(1+\beta_1)}{2\Sigma} \right)
\end{equation}

Note that in Equation~\ref{eqn:pbal} we have
just considered the ram pressure in the upstream.
We have neglected gas pressure, since the shock velocity is highly supersonic.
We have also not included magnetic pressure in the pre-shock region, which is in equipartition with the ram pressure with a ratio
\begin{equation}
R = \frac{B_1^2/(8\pi)}{\rho_1 v_1^2} 
= \frac{  \rho_1 c_{\rm s}^2/\beta_1 }{\rho_1\mathcal{M}^2 c_{\rm s}^2} 
=\frac{1}{\beta_1\mathcal{M}^2}
= \frac{\rho_1^2}{\rho_0^2 + \rho_0\rho_1(1+\beta_1)}
\end{equation}

Fig.~\ref{fig:1dprofile}
shows the evolution of the 1D shock problem
for the case of a supersonic hydrodynamic shock and the 
case of a parallel magnetized shock with $\beta_{{\rm init},\perp} = 0.25$.
When the hydrodynamic shock breaks out, it quickly achieves a self-similar exponential profile. 
In the case of the parallel magnetized shock, 
the shock propagates much faster than the hydrodynamic shock owing to the fast magnetosonic mode. It also has a smaller density contrast but sweeps up more mass. The initial shock profile is not exponential due to the additional magnetic pressure support (and resembles the analytic model plotted in Fig.~\ref{fig:1danalytic}), but the solution soon evolves into an exponential atmosphere solution as peak density drops since the magnetic pressure decreases with density and the scenario resembles 
a pure hydrodynamic case except for the fact that the shock continues to travel above the fast magnetosonic speed.

We show the evolution of the mass per unit area of the shock, the peak density, the sonic Mach number, and the scale height evaluated as $h=\Sigma/\rho_0$ in
Fig.~\ref{fig:1dx4} for 5 different shocks characterized by
$\beta_{{\rm init},\perp}=\infty, 25,0.25,0.028,0.0025$.
The effect of the magnetic field for a shock traveling perpendicularly to it is to increase the mass in the shock, greatly decrease the density contrast, increase the Mach number, and increase the effective scale height.

The analysis provides the insight that in a strongly magnetized supersonic fluid, 
most of the high density structures will be oriented parallel to the magnetic field.
This organization of structure with strong magnetic fields is seen in 3D turbulent simulations, 
including the simulation suite from \cite{2017ApJ...838...40M} which we analyze next in Section~\ref{sec:3d} (see also the visualization of density structures and the magnetic field in (Fig.~\ref{fig:sims})).
The anisotropy has important implications for the geometry under which stars form.
Stirring the gas in directions perpendicular to the field lead to shocks with greatly reduced peak densities that disperse at much faster speeds. For a dense region to collapse under gravity, the gravitational free-fall time must be shorter than the shock expansion timescale, which would mean it is more difficult for perpendicular shocks to collapse under self-gravity.  We explore this further in the next section.

\begin{figure}
\begin{center}
\begin{tabular}{ccc}
\rotatebox{90}{\qquad$\beta_{\rm mean-field}=25$} &
\includegraphics[width=0.39\textwidth]{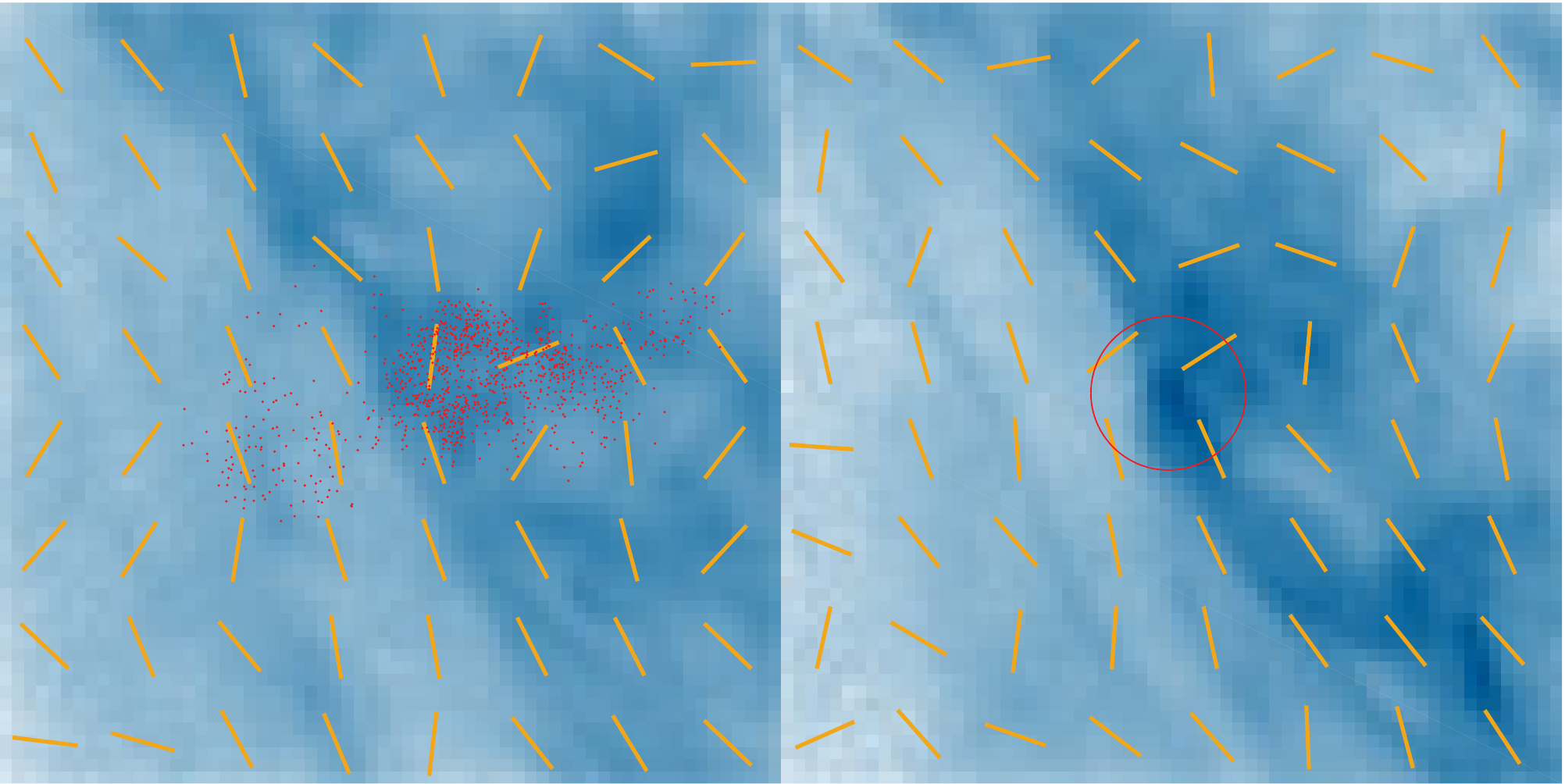} & \multirow{4}{*}{\includegraphics[width=0.05\textwidth]{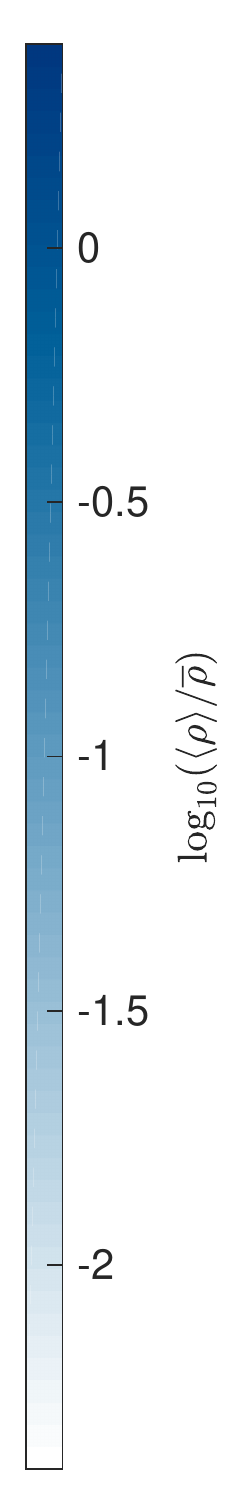}} \\
\rotatebox{90}{\,\quad$\beta_{\rm mean-field}=0.25$} &
\includegraphics[width=0.39\textwidth]{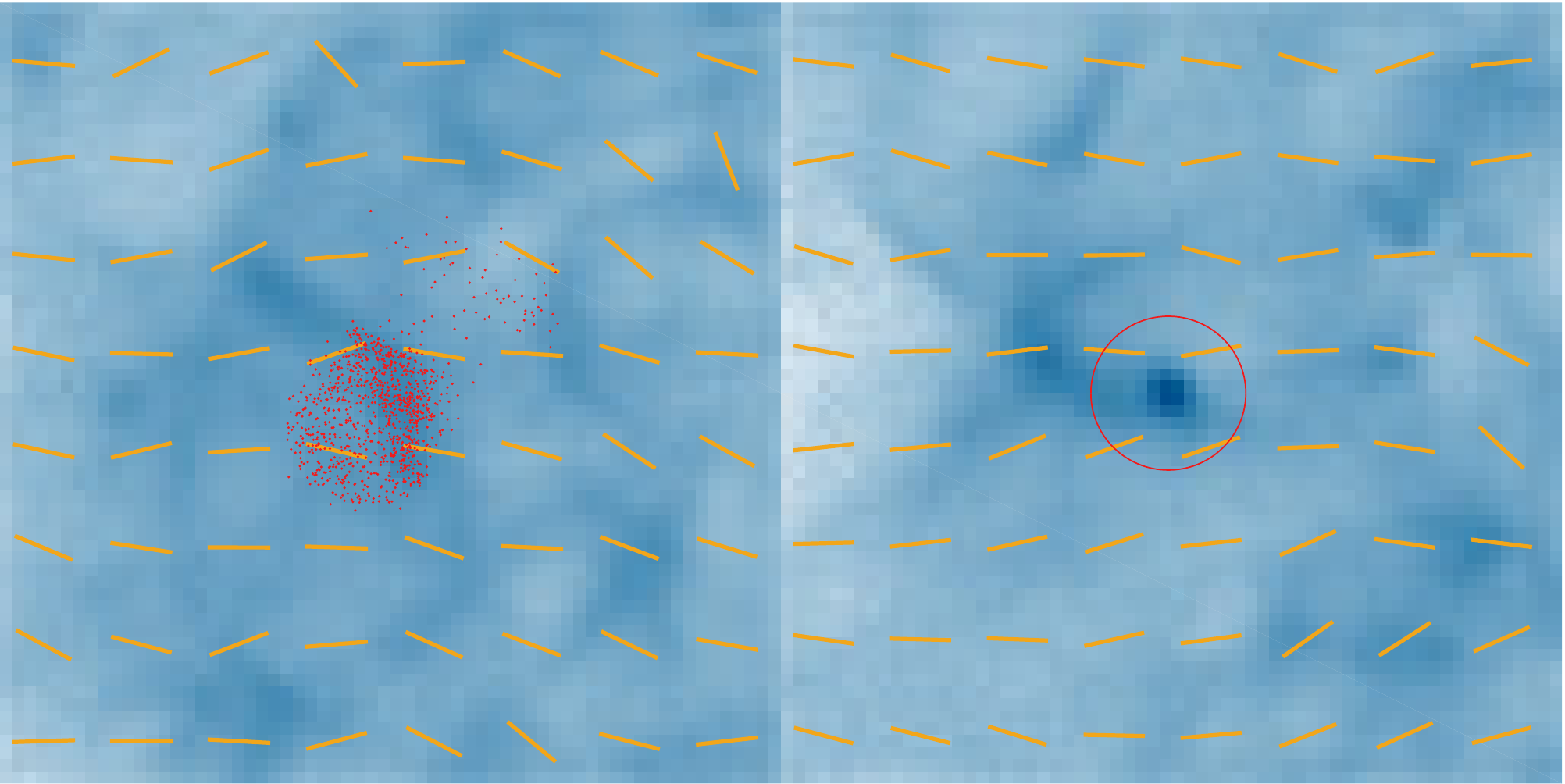} & \\
\rotatebox{90}{\,\,\,\,\,\,\,$\beta_{\rm mean-field}=0.028$} &
\includegraphics[width=0.39\textwidth]{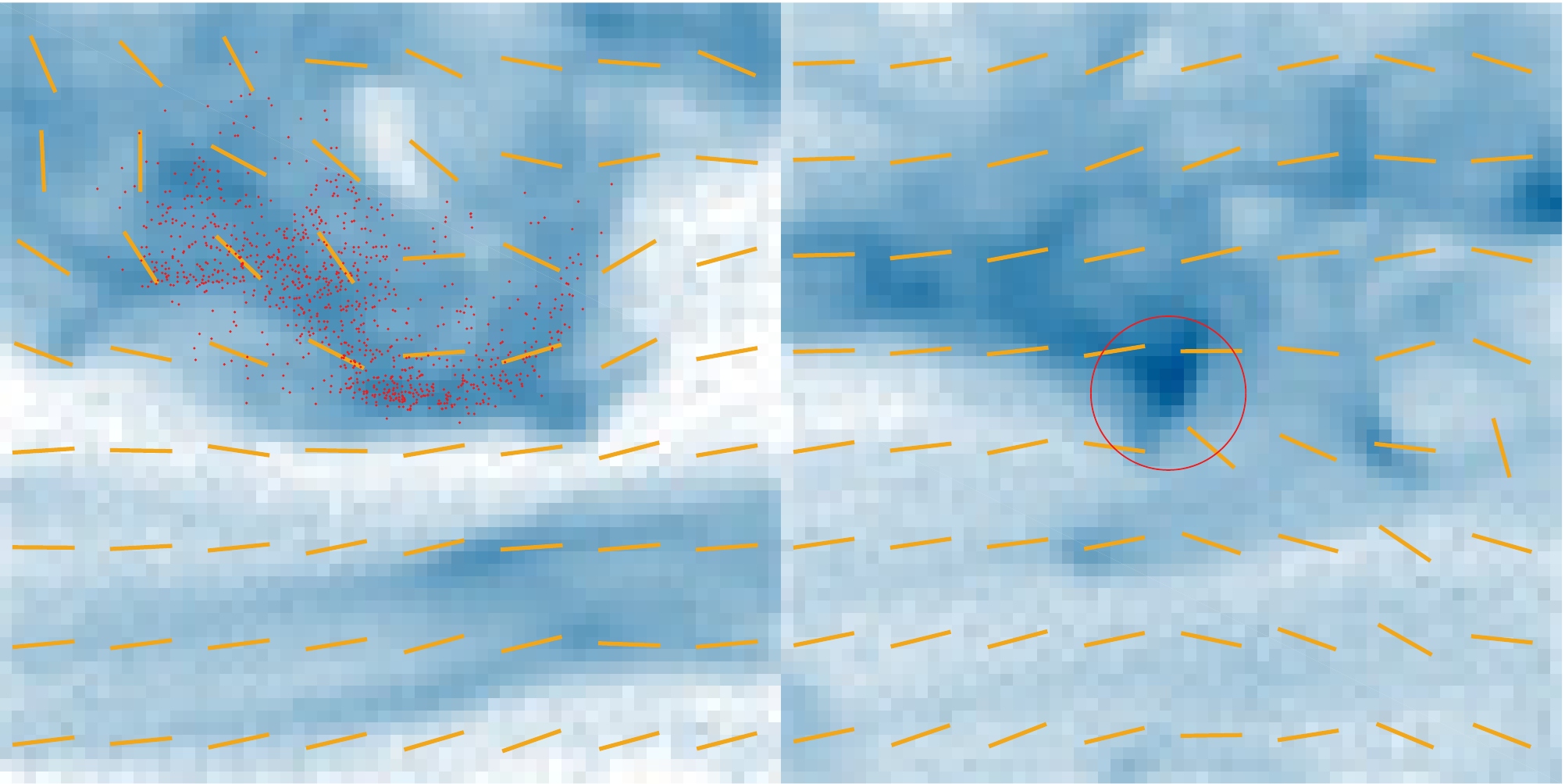} & \\
\rotatebox{90}{\,\,\,\,\,\,\,$\beta_{\rm mean-field}=0.0025$} &
\includegraphics[width=0.39\textwidth]{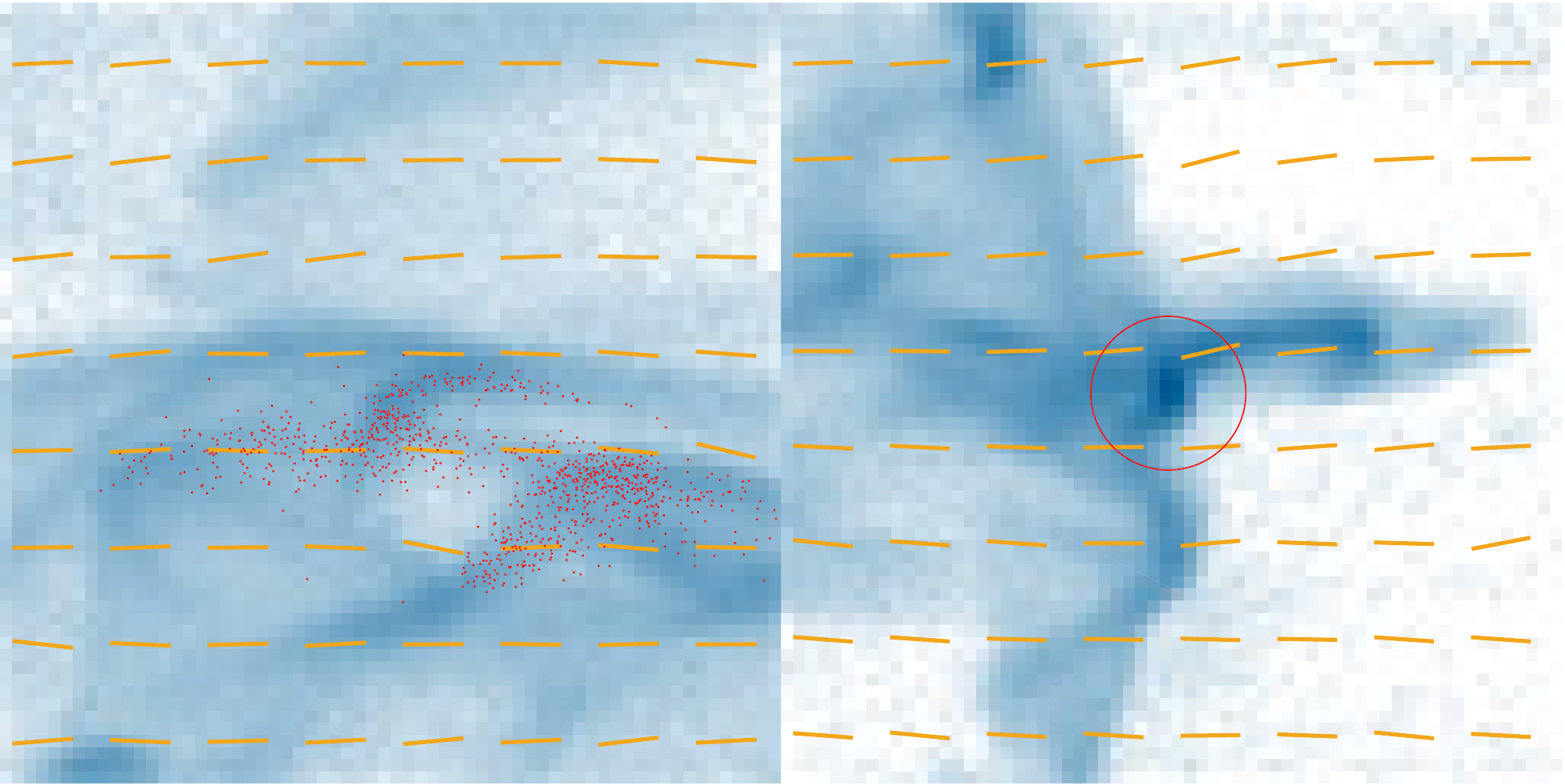} &
\end{tabular}
\end{center}
\caption{Projected densities zoomed in on a $L/10\times L/10$ region centered around a collapsing core: precollapse (left column) and collapse (right column) for 4 simulations with varying different mean-field strengths. Shown also are polarization vectors for the magnetic field and the original location of tracer particles (red dots) that have ended up in the collapse core within radius $L/100$ (red circle).}
\label{fig:proj}
\end{figure}

\begin{figure}
\begin{center}
\begin{tabular}{cc}
\rotatebox{90}{\,\,\,\quad\qquad$\beta_{\rm mean-field}=25$} &
\includegraphics[width=0.32\textwidth]{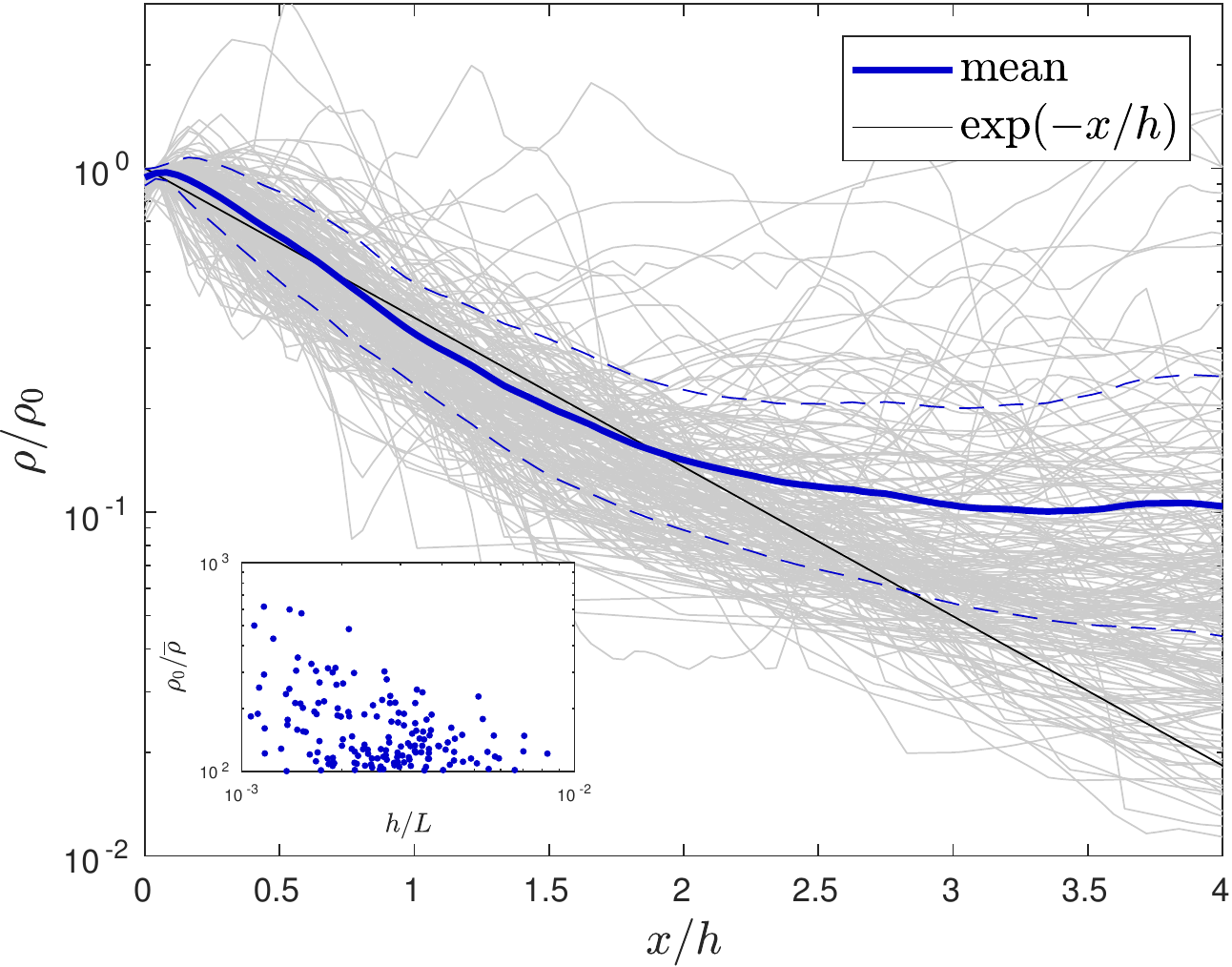} \\
\rotatebox{90}{\,\qquad\quad$\beta_{\rm mean-field}=0.25$} &
\includegraphics[width=0.32\textwidth]{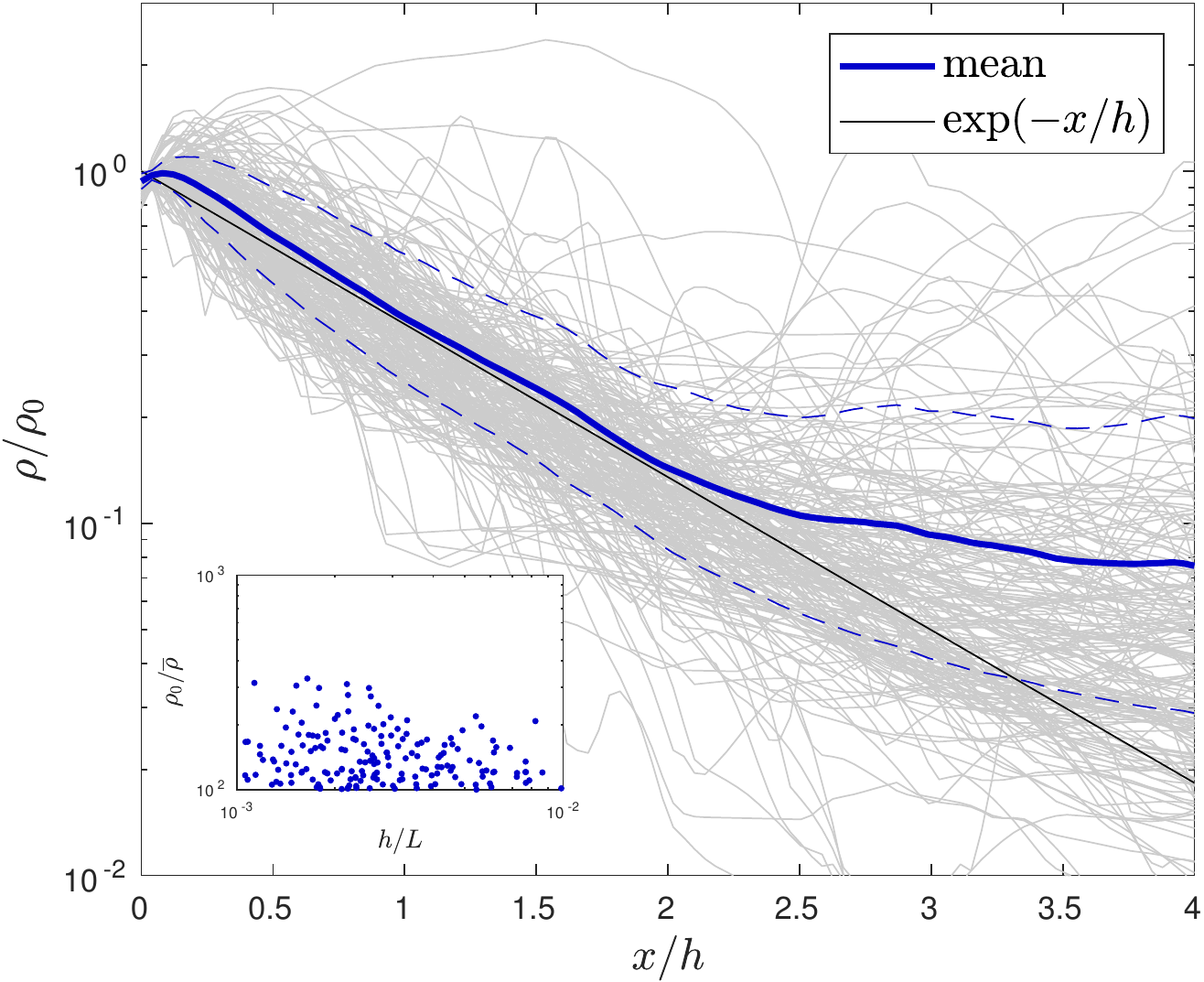} \\
\rotatebox{90}{\,\,\,\,\,\qquad$\beta_{\rm mean-field}=0.028$} &
\includegraphics[width=0.32\textwidth]{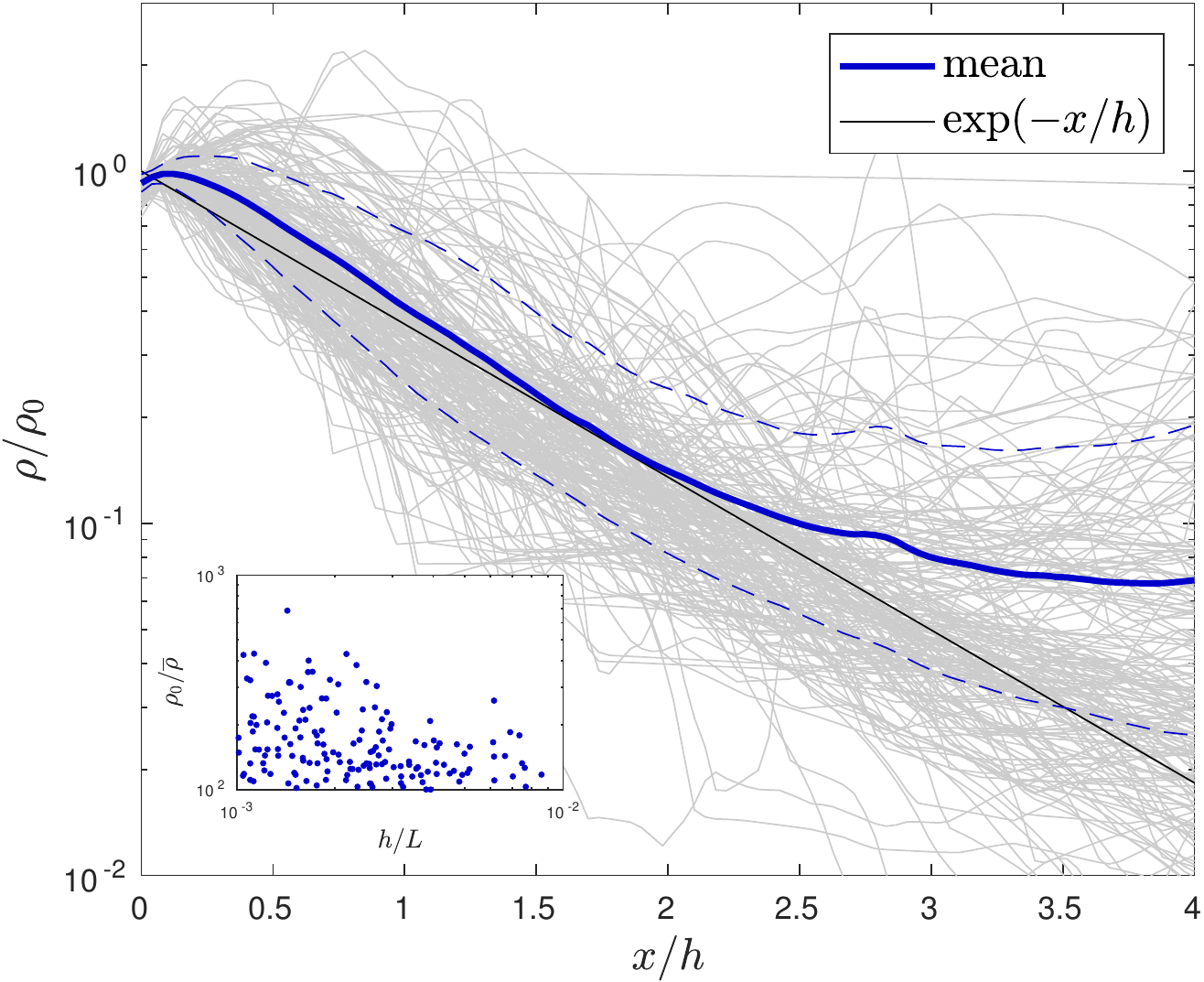} \\
\rotatebox{90}{\,\,\,\,\qquad$\beta_{\rm mean-field}=0.0025$} &
\includegraphics[width=0.32\textwidth]{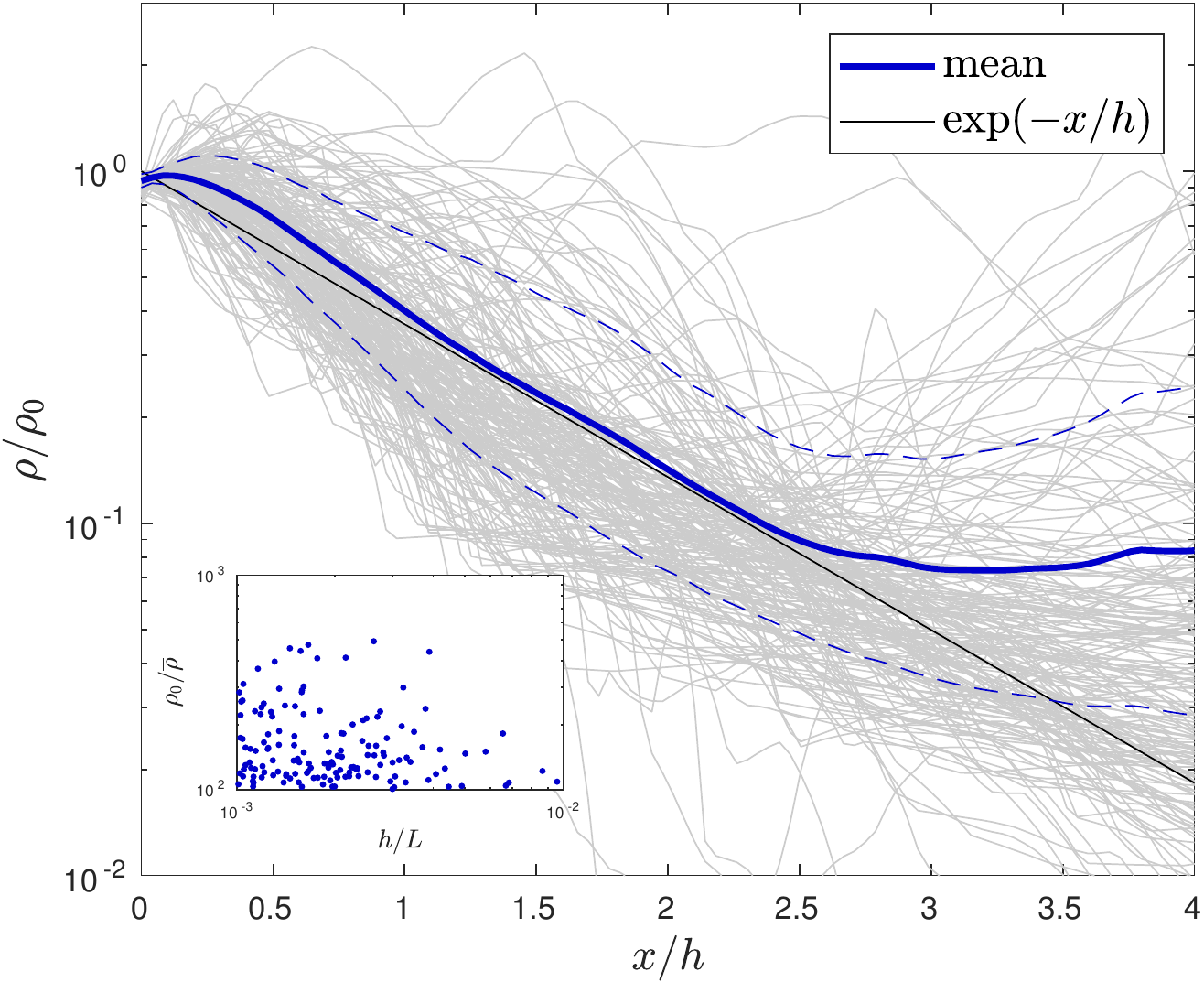}
\end{tabular}
\end{center}
\caption{Post-shock density profiles of shocked regions in the simulations of supersonic isothermal turbulence. A path through each individual shocked region oriented with the shock was drawn, and the profile was scaled by the best-fit scale height. The average-profile (solid blue line) resembles an exponential atmosphere. The standard deviation of the distribution is also shown as dashed lines. No strong dependency of the average profile is found with magnetic field strength, although the exponential profile extends over slightly more scale-heights in the strong field case. The inset shows the distribution of the characteristic sizes and densities of these shocks.}
\label{fig:shocks}
\end{figure}

\begin{figure*}
\begin{center}
\begin{tabular}{ccc}
\rotatebox{90}{\qquad\qquad\quad$\beta_{\rm mean-field}=25$} &
\includegraphics[width=0.3\textwidth]{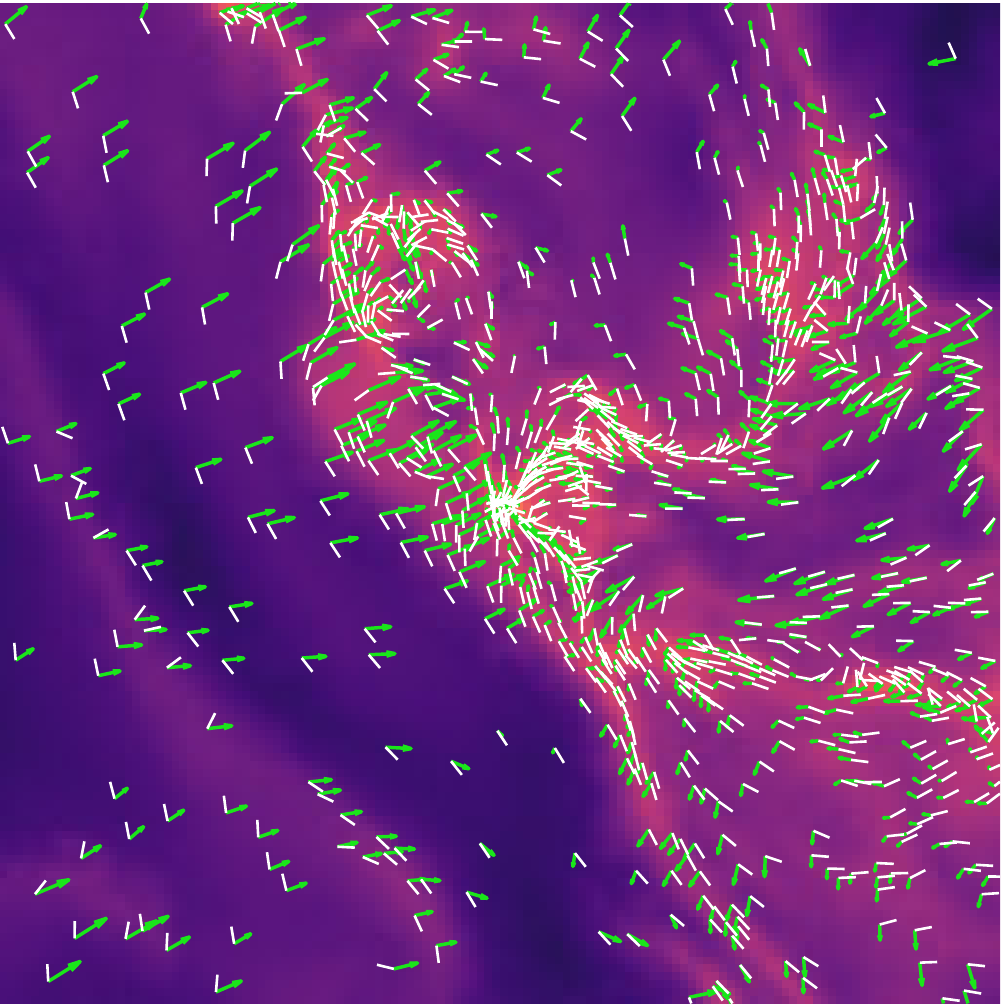}
\includegraphics[width=0.3\textwidth]{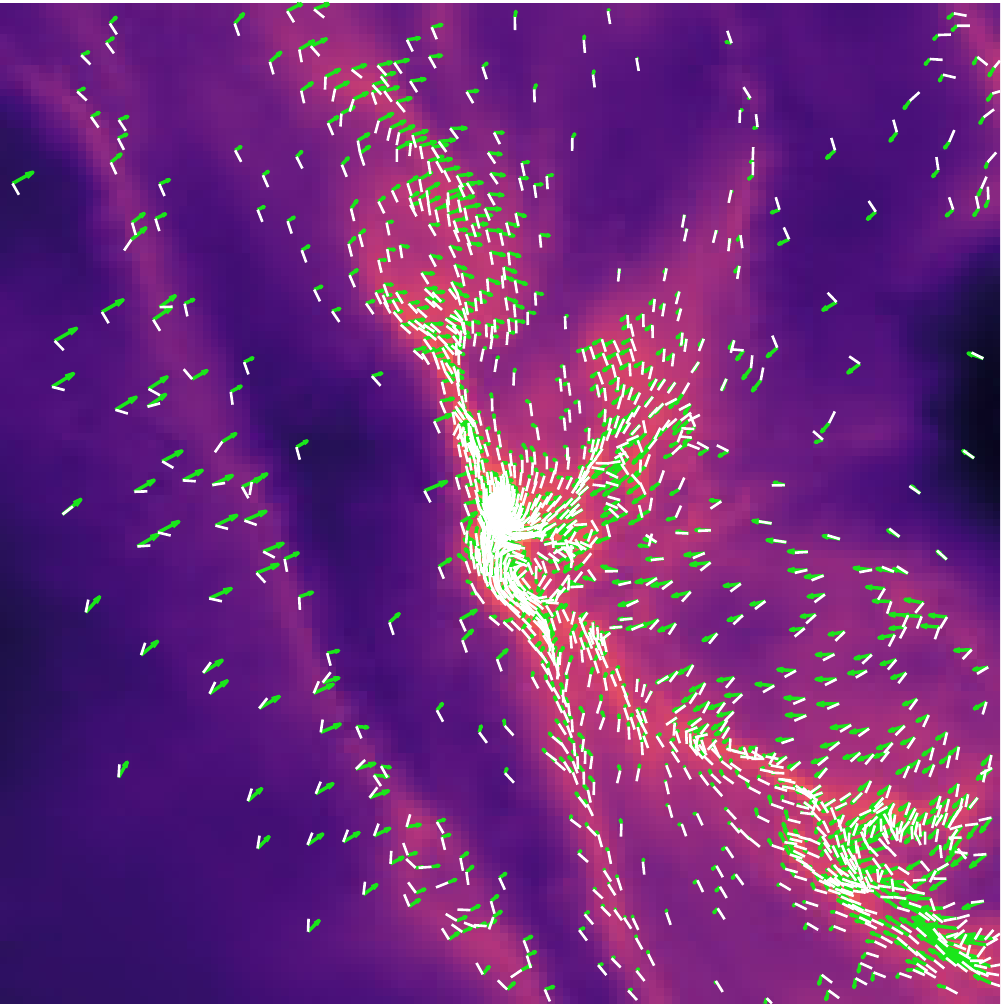} & \multirow{4}{*}{\includegraphics[width=0.1\textwidth]{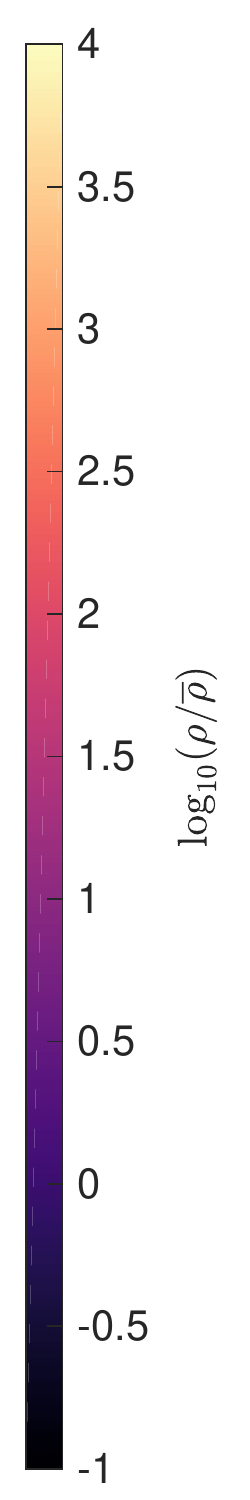}} \\
\rotatebox{90}{\qquad\qquad$\beta_{\rm mean-field}=0.25$} &
\includegraphics[width=0.3\textwidth]{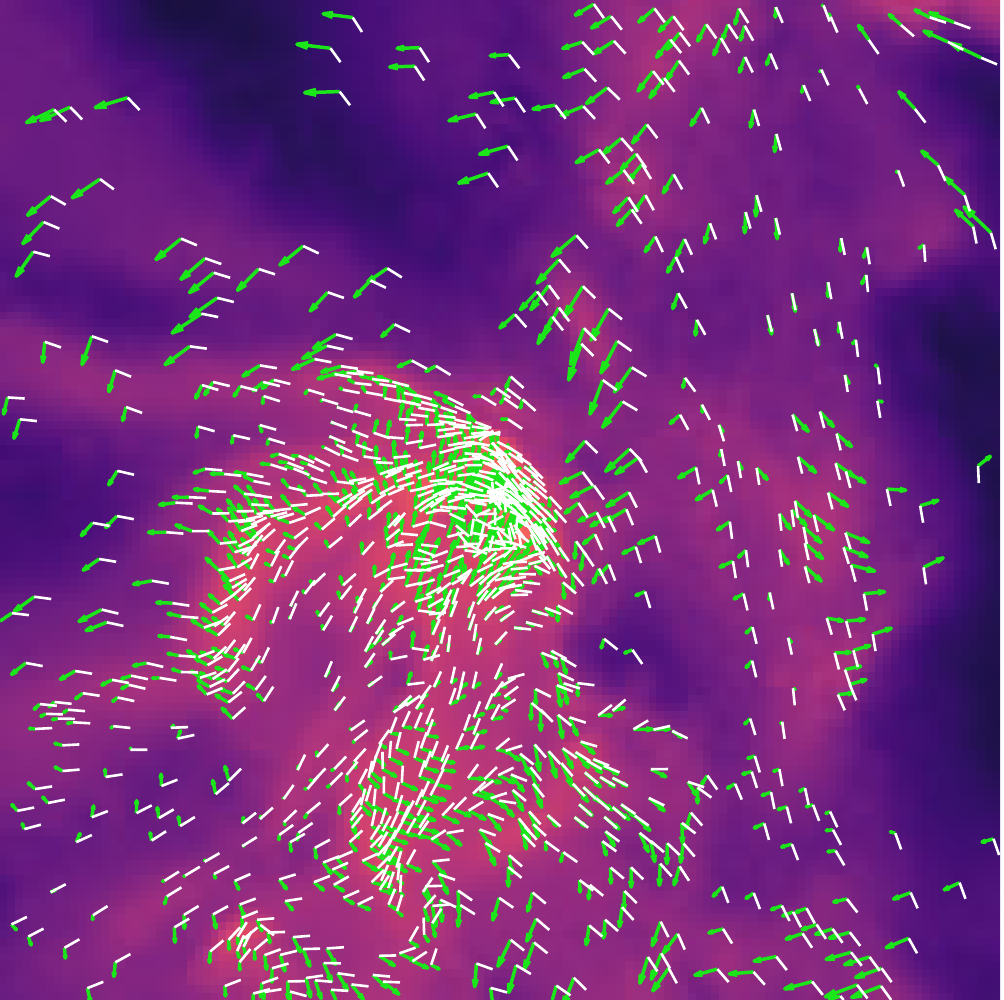}
\includegraphics[width=0.3\textwidth]{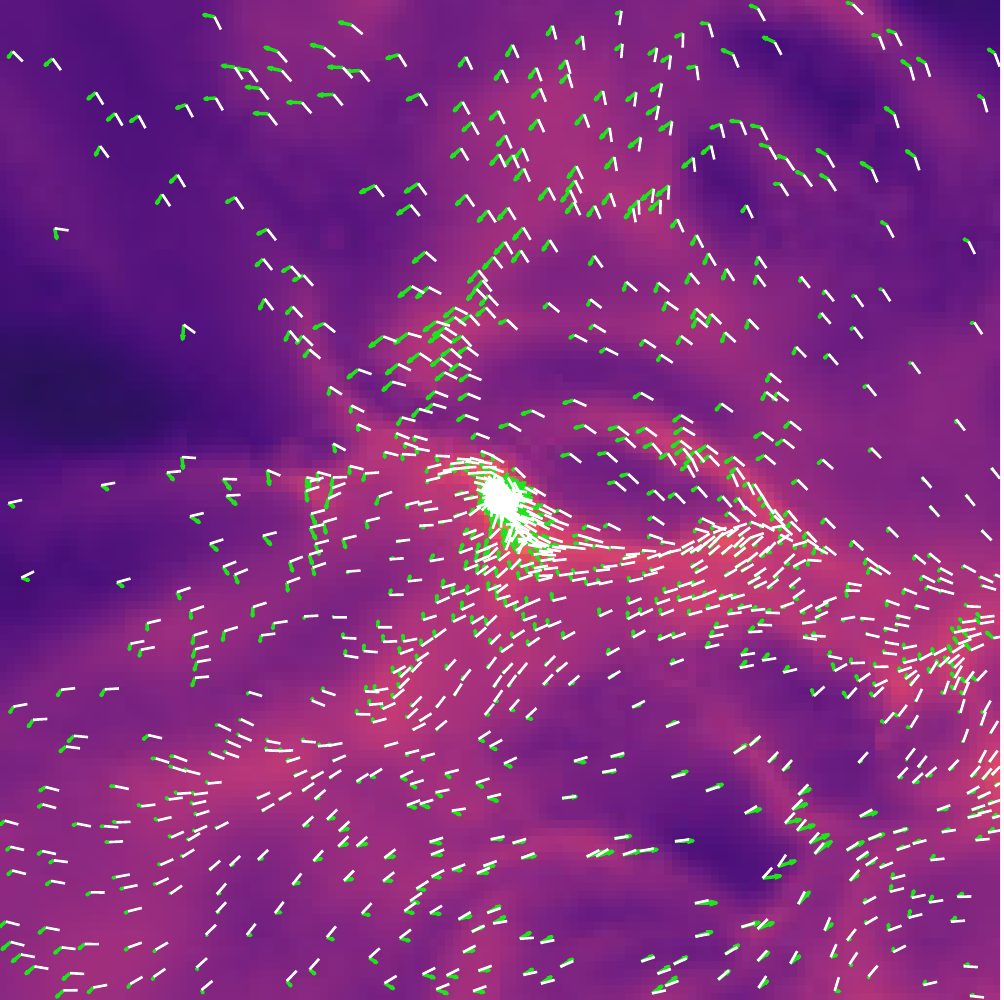} &\\
\rotatebox{90}{\qquad\qquad$\beta_{\rm mean-field}=0.028$} &
\includegraphics[width=0.3\textwidth]{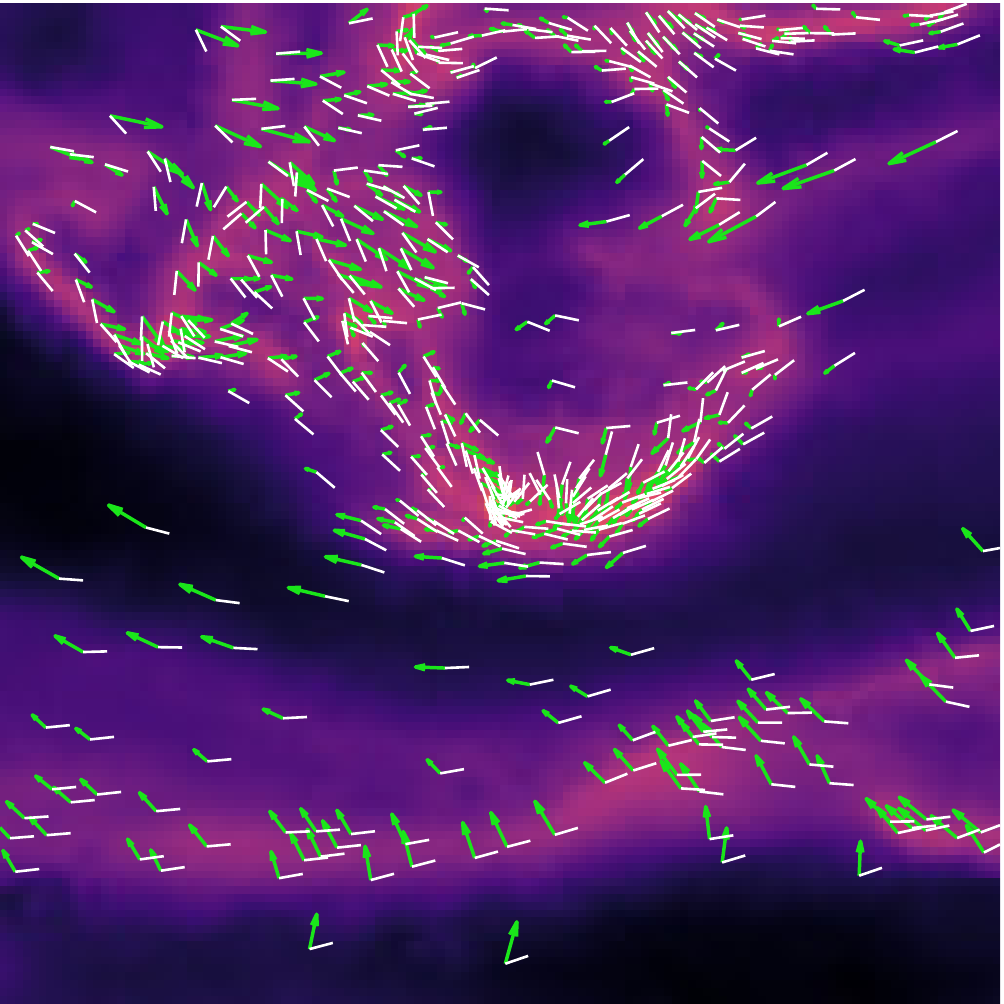}
\includegraphics[width=0.3\textwidth]{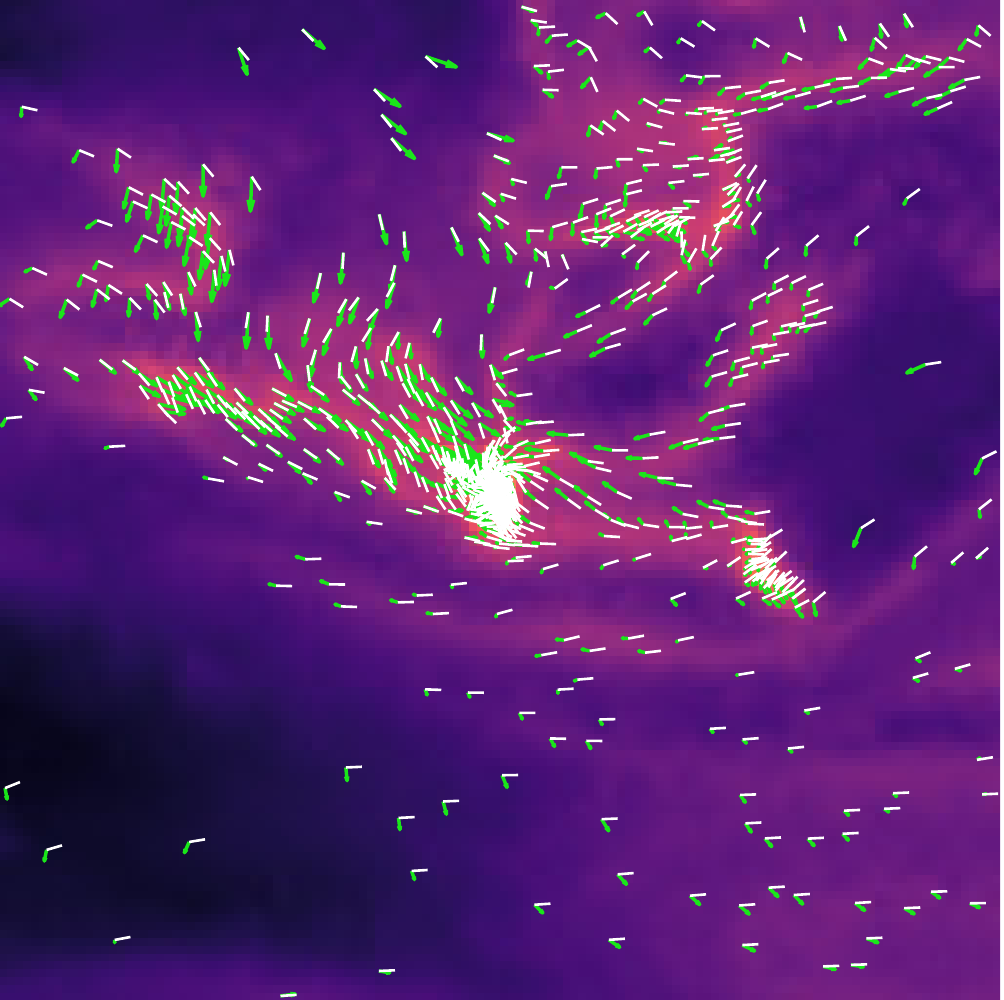} & \\
\rotatebox{90}{\,\qquad\quad$\beta_{\rm mean-field}=0.0025$} &
\includegraphics[width=0.3\textwidth]{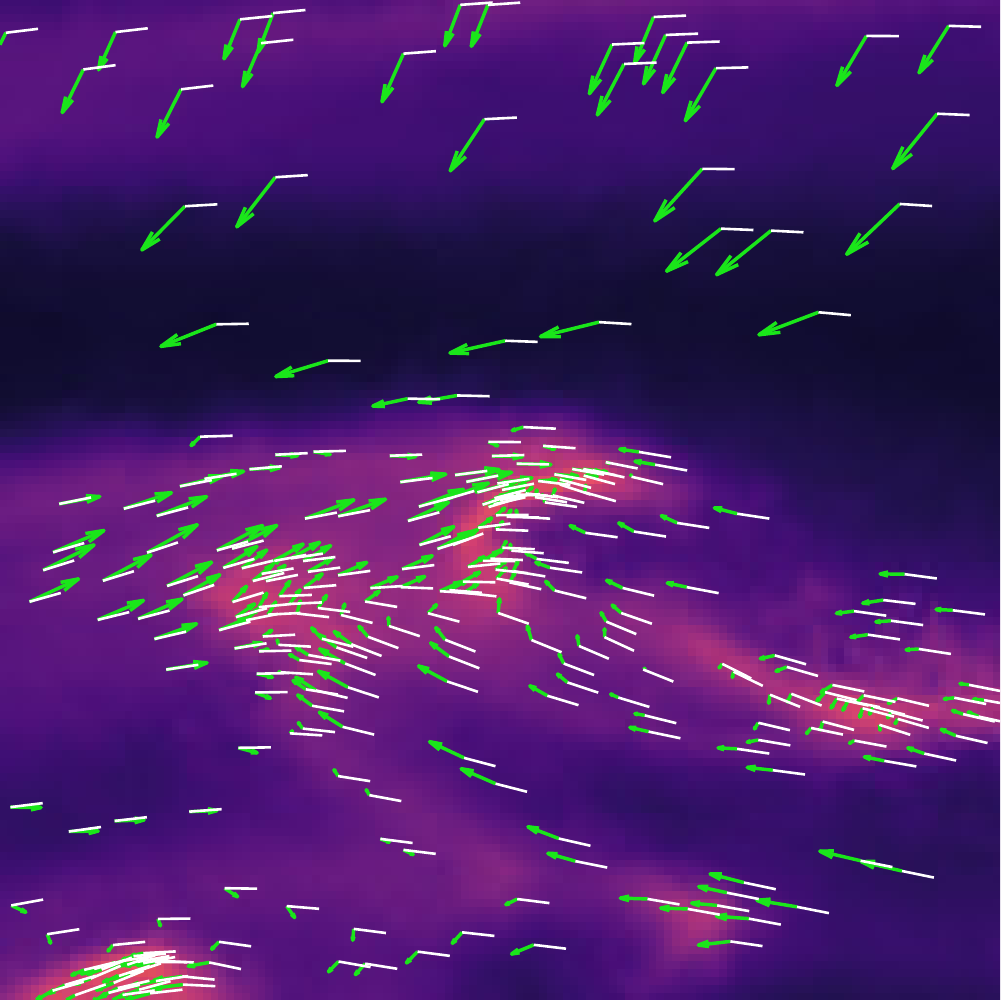}
\includegraphics[width=0.3\textwidth]{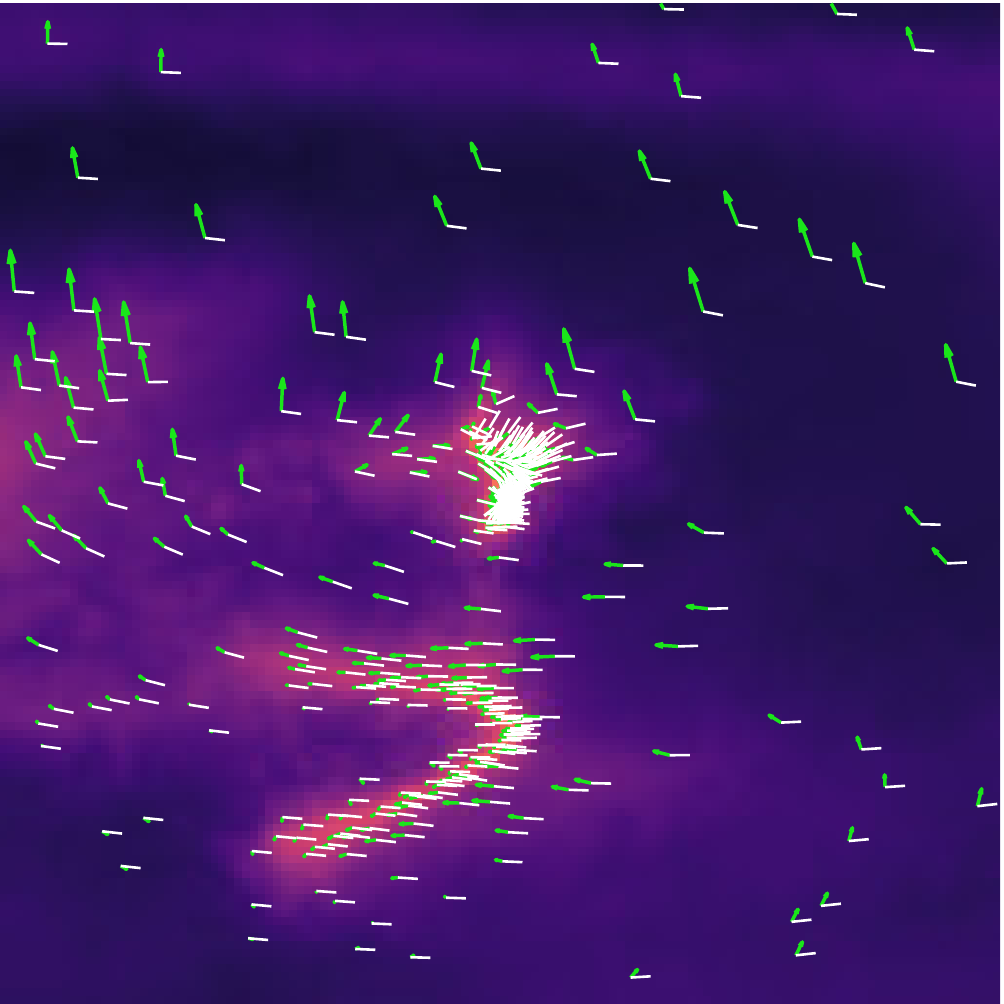} &
\end{tabular}
\end{center}
\caption{Density slices zoomed in on a $L/10\times L/10$ region centered around a collapsing core: precollapse (left column) and collapse (right column) for 4 simulations with varying different mean-field strengths. Shown also are polarization vectors (white) for the magnetic field and arrowed vectors (green) for the velocity of Lagrangian particles.}
\label{fig:slice}
\end{figure*}

\begin{figure*}
\begin{center}
\begin{tabular*}{0.65\textwidth}{c @{\extracolsep{\fill}} c}
$\beta_{\rm mean-field}=0.0025$ & $\beta_{\rm mean-field}=25$
\end{tabular*}\\
\includegraphics[width=0.47\textwidth]{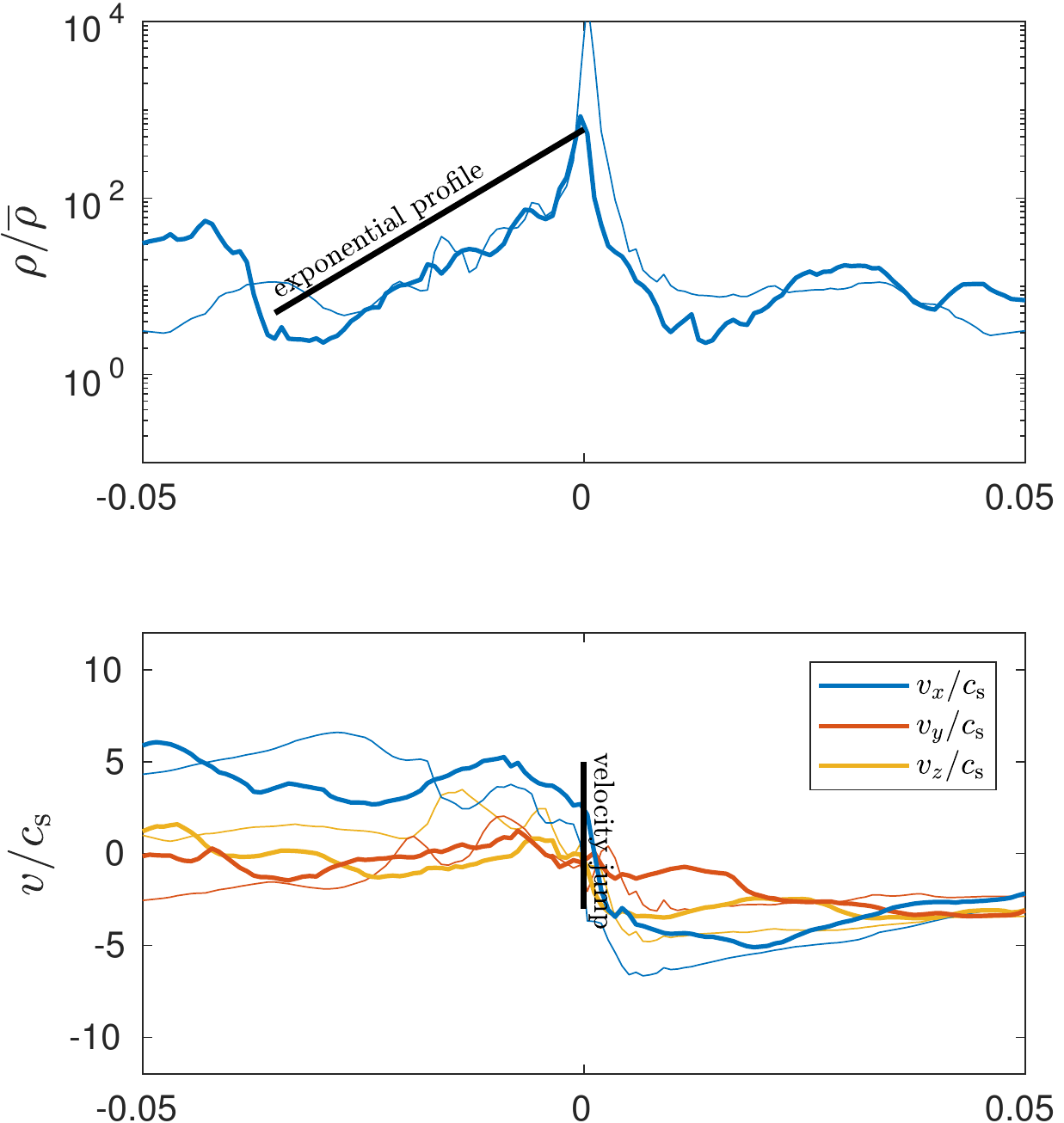}
\includegraphics[width=0.47\textwidth]{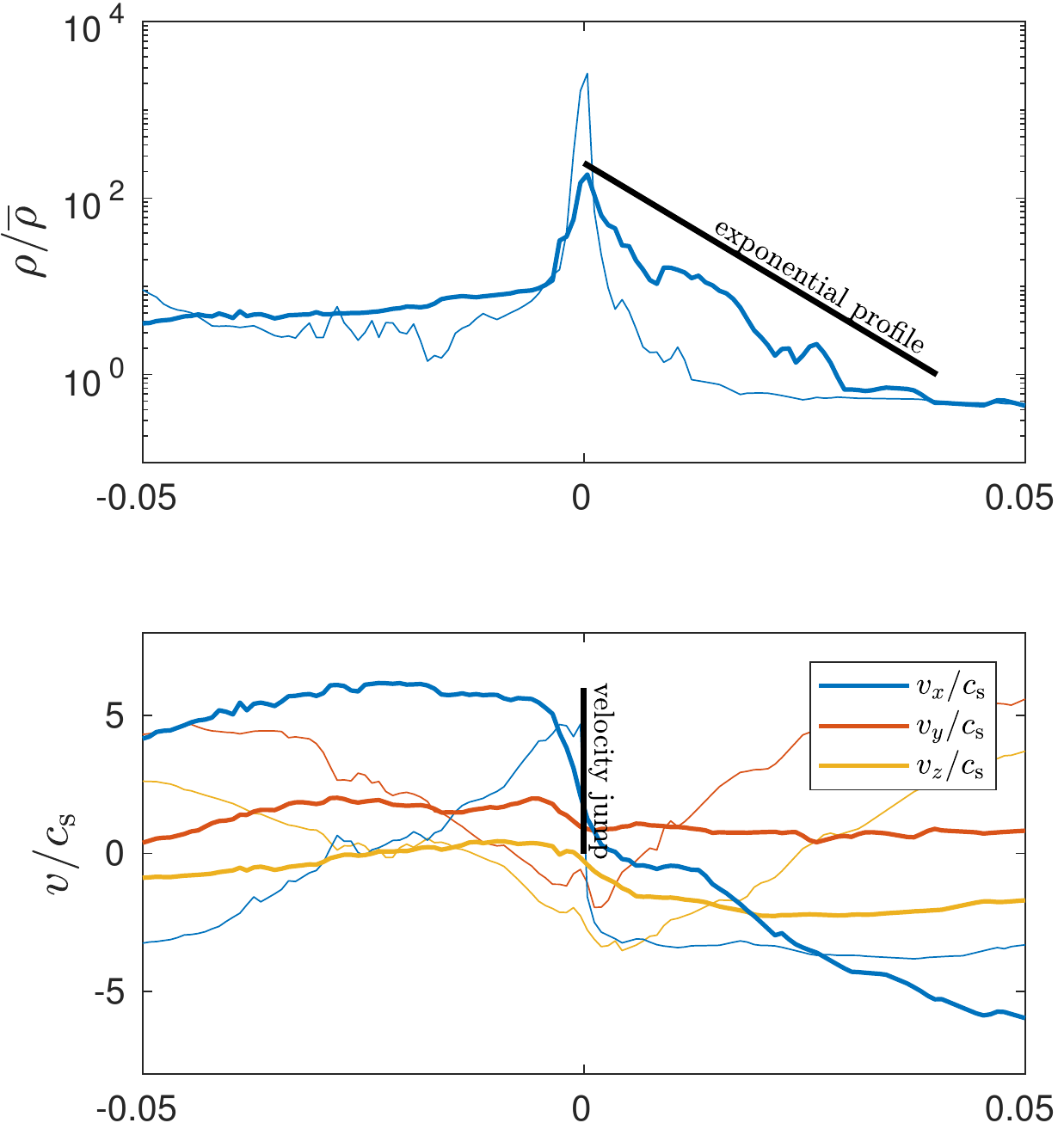}
\end{center}
\caption{Examples of dense shock structures identified before collapse (thick lines) which become collapsed pre-stellar cores (thin lines), shown for simulations with two different magnetic mean-field strengths (weak and strong).
Prior to collapse, an exponential density structure is evident, with discontinuity in the parallel velocity field.}
\label{fig:s1d}
\end{figure*}

\section{Gravitational collapse in 3D isothermal magneto-turbulence}
\label{sec:3d}

Here we consider the origins and collapse of dense structures in supersonic, isothermal turbulence under self-gravity.
The details of the simulations have been described in Section~\ref{sec:sims}.
We track the collapse of the first overdensities from few parsec to $<100$~AU scales. The simulations are meant to capture the collapse of gas in a GMC and form a pre-stellar core at the end of its isothermal collapse phase.
We are primarily interested in the origins and initial geometry of the collapse here. We have explored the properties of the cores that ultimately form in \cite{2017ApJ...838...40M}.

Projections of pre-collapsed and collapsed gas (after $\sim 0.5 \tau_{\rm free-fall}$) are shown in Fig.~\ref{fig:proj}, to illustrate a sense of the types of structures being considered here. The figure only shows a region of boxsize $L/10$ centered around 
the densest point that undergoes collapse. The fluid elements that make up a collapsed core can be traced back to their origins prior to collapse (e.g. red points in the left panels of Fig.~\ref{fig:proj}), 
and are found to arise from density peaks in strong post-shock regions. The figure also shows the line-of-sight density-averaged magnetic field vectors (yellow lines), and red points that trace the origins of the collapsed gas within a radius of $L/100$. 
Note that the site of star formation does not stay fixed, we have recentered the projections on the densest originating gas cell. The location of the core can move significantly with the large-scale eddy turnover time.
The tracers indicate that in the case of a strong magnetic field (i.e. the bottom two panels of Fig.~\ref{fig:proj}) the overdense regions that collapse into a core are elongated along the magnetic field lines, the direction which does not experience magnetic tension.
Properties of such collapsing cores (energy densities, radial profiles, magnetic field morphologies) have been analyzed extensively in \citet{2017ApJ...838...40M}, but their origins have not been traced.

We identify dense-regions ($\rho>100\overline{\rho}$) in the isothermal simulations prior to gravitational collapse and draw a path oriented with the density gradients. The resulting profile is then fit with an exponential atmosphere profile. Individual scaled profiles and their average is shown in Fig.~\ref{fig:shocks}. The exponential model is a good fit to the dense post-shock regions, at least over two scale heights, even in the case of a strong magnetic field threading the domain. This result is a validation of the results of \cite{2018arXiv180105440R} and an extension to the case of MHD turbulence. The exponential profile extends over a slightly larger range of scale heights as the mean-magnetic field is increased, which we attribute to the fact that the magnetic field helps orient the shocks (parallel shocks have the largest density contrast and survive the longest). The conceptual picture of \cite{2018arXiv180105440R} is hence even clearer in a strongly magnetized medium due to the anisotropy and order the large scale field creates. The mixing motions of turbulence in this case are more limited to perpendicular to the magnetic field lines. Fig.~\ref{fig:shocks} also shows the distribution of the characteristic sizes and densities of the shocked sheets in the inset, which are largely unaffected by the mean magnetic-field threading the box. Shocked sheets break out with large density peaks, which decrease as they grow in size.
The shocks have surface densities of characteristic value 
$\Sigma \equiv \rho_0 h \lesssim \overline{\rho}L$.

We show slices of density centered on the pre-collapsed and collapsed gas in Fig.~\ref{fig:slice}.
This highlights some of the structure in the gas that can be difficult to see in a projection (e.g. shock fronts may not be apparent in projections).
Additionally, the velocity and magnetic field direction of the quasi-Lagrangian (roughly equal mass) gas cells in the simulation are shown. What is clear from the figure panels is that the cores form in a network of shocks, which can be seen to be sweeping up unshocked material (we can see a discontinuous velocity field ramming into each other). The morphology of the converging flow is sheetlike. Converging flows are also seen towards the gravitationally collapsed core centers (as shown with green arrows).

To further study the shocked gas, we identify and show line-of-sight cuts in a direction approximately parallel to the shock direction
in Fig.~\ref{fig:s1d}.
Pre-collapsed (thick lines) and collapsed (thin lines) profiles are compared. The pre-collapse profiles show strong evidence of an exponential atmosphere structure, aligned with the magnetic field direction in the case that the field is strong, as predicted by the 1D models. Velocities across the shock interface are discontinuous and indicate pre-shock material is being swept up across the shock front. The pre-collapse density peaks are approximately $\mathcal{M}^2$ in strength, where $\mathcal{M}$ is the shock Mach number which can be measured from the jump in $v_x$. Collapsed structure becomes strongly core-like but signatures of the shock still remain after a fraction of the free-fall time on larger scales.

Importantly, our simulations highlight the regions in isothermal turbulence that ultimately evolve into pre-stellar cores, and give descriptions of the properties of the initial configurations from which they grow. Star formation in the context of this model is found to be strongly associated with strong shocks. In the case of a strong large-scale magnetic field (sub-Alfvenic turbulence), these shocks are also oriented parallel to the magnetic field, with dense structures forming perpendicular to the mean field, similar to what has been observed by \textit{Planck} \citep{2016A&A...586A.138P,2017A&A...607A...2S}. These shocks, the regions that do not collapse, disperse on the order of the free-fall time, so they would only be seen to be clearly spatially associated with young protostars.

\begin{figure}
\begin{center}
\includegraphics[width=0.47\textwidth]{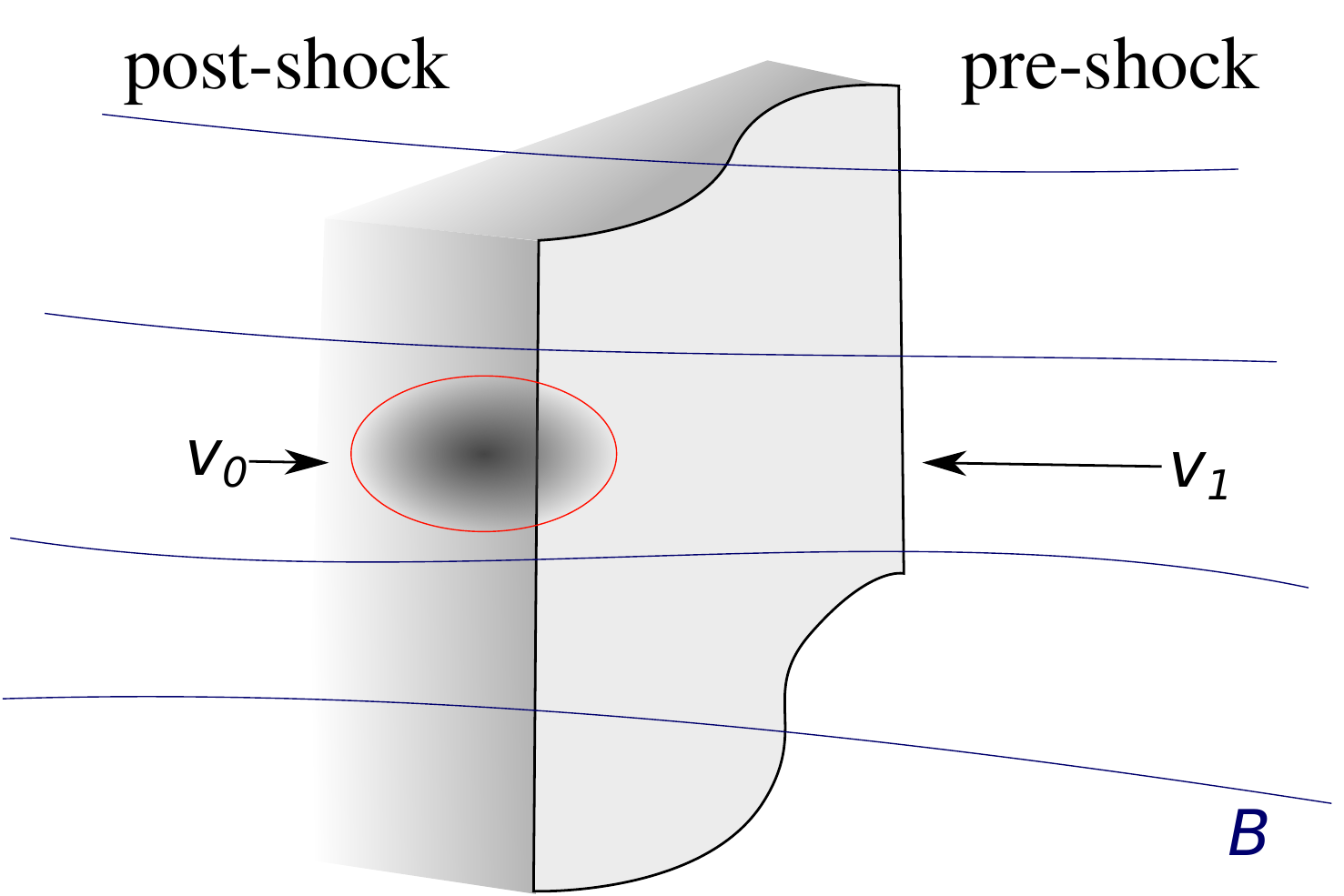}
\end{center}
\caption{A schematic illustrating the process of pre-stellar core formation in a shocked region of supersonic turbulence. The post-shock region, preferentially traveling parallel to a strong magnetic field, sweeps up
shock material and shows an exponential density profile and may undergo Jeans instability to collapse into a core (red outline). In the case the magnetic field is strong, the collapse in the perpendicular directions may need to be mediated by a magnetic flux removal process such as fast reconnection diffusion \citep{LEC12}. Perturbations in the overdensities are also preferentially elongated along a strong magnetic field line.}
\label{fig:diagram}
\end{figure}

\section{Discussion}
\label{sec:disc}

A schematic illustrating the process of pre-stellar core formation from shocked regions of supersonic MHD turbulence is shown in Fig.~\ref{fig:diagram}. The dense post-shock region, characterized by an exponential profile and traveling parallel to strong mean magnetic fields, may undergo gravitational collapse. 

In the case that the magnetic field is not strong, these shocks are oriented randomly in the fluid. As pointed out by \cite{2018arXiv180105440R}, the volume-filling fraction of these shocks are small, so interaction between two highly shocked regions is rare, and simple exponential profiles of shocks can be easily identified in a turbulent simulation. For collapse to occur, the free-fall timescale needs to be less than the expansion timescale for these dense regions.
 We have demonstrated that indeed these shocked sites are the origins of collapsing cores in our numerical simulations with self-gravity.

In the case of a strong magnetic field, shocks traveling perpendicularly to the magnetic field have reduced density contrast and significantly faster dispersion timescales owing to the fast magnetosonic speed (Section~\ref{sec:1d}).
Shocks in the parallel direction behave identically to the no magnetic field case, achieving a large density contrast of $\mathcal{M}^2$ and expanding with the sound speed. Thus these parallel shocks are more likely to collapse, as we indeed see in our simulations.
High density pre-collapse structures are aligned parallel with the magnetic field, owing to the  perpendicular mixing motions of turbulence and shock compression \citep{2018arXiv180200987X}.

In the case that the magnetic field is very strong, the flux-frozen condition (initial mass-to-flux ratio) will prevent collapse perpendicular to a large-scale magnetic field that threads the fluid under the assumptions of ideal MHD. 
This would be indeed the case with our strongest magnetic field simulation ($\beta_{\rm mean-field}=0.0025$ ). However, as we found in \cite{2017ApJ...838...40M}, 
there is actually a loss of magnetic flux in the core as it collapses in the simulations, which we have attributed to fast turbulent reconnection diffusion
\cite{1999ApJ...517..700L,2003LNP...614..376V,2012ApJ...747...21S,2004ApJ...603..180L,LEC12,2013ApJ...777...46L}. Such a naturally occurring process in turbulence is an effective way to lose magnetic flux through the reconnection of field lines \citep{2017ApJ...838...91K}. Importantly, the process is independent of resistivity \cite{2009ApJ...700...63K} in this regime of turbulence. Despite the fact that we simulated turbulence with the ideal MHD assumption, small numerical resistivity due to truncation errors is present in our simulations, which enabled fast magnetic reconnection (and consequently collapse) to occur.
We see clear evidence of reconnection morphology in 
the sliced image for the $\beta_{\rm mean-field}=0.0025$ core (Fig.~\ref{fig:proj}, just north of the core center), despite the fact that this is not evident in the projected magnetic field, which looks much simpler: a classical pinched hourglass shape (Fig.~\ref{fig:slice}).

The physical, conceptual picture that arises in the case of a strong magnetic field from our simulations is that collapse along field lines is uninhibited and will create shocks moving in a direction aligned with the large-scale magnetic field, with density enhancements piling up perpendicular to the magnetic field (e.g. see Figure 4). This picture explains the observed orientation of density gradients and magnetic field and may suggest much of the diffuse ISM is sub-Alfv\'enic \citep{2016A&A...586A.138P,2017A&A...607A...2S}.

However, an additional physical process may be needed to help aid collapse in the perpendicular direction to overcome the magnetic tension. This may by fast reconnection diffusion, as in our simulations. But non-ideal MHD processes may also be invoked, such as
ambipolar diffusion, which would act on a longer timescale to expel the magnetic field and contract \citep{2014ApJ...785...69C}. In our simulations, we do not include ambipolar diffusion effects, yet nevertheless collapse occurs in the strong field case.

In \cite{2017ApJ...838...40M}  we have identified two modes of pre-stellar core collapse based on the collapsed core properties (e.g. how the magnetic field grows as a function of density) in regimes where either the turbulent kinetic energy or the mean magnetic field dominates over the other. The two regimes are (1) isotropic, non-self-similar ($\beta$ decreases) collapse (weak-field) and  (2) anisotropic, self-similar ($\beta\sim 1$) collapse (strong-field). This picture is further supported by the dichotomy seen in the originating shocks that undergo collapse. The main difference that causes these two modes of collapse is the additional overcoming of the magnetic tension perpendicular to collapse in the strong-field case.

\cite{2018arXiv180105440R} shows that dense shocks make up nearly $100$ per cent of the dense gas in an isothermal supersonic medium, constituting the high-density end of the log-normal distribution of gas densities at $\rho\gtrsim 25\overline{\rho}$.
A number of works have shown that under gravitational collapse, a powerlaw tail in the probability distribution function (PDF) at high densities develops, which scales as $\rho^{-1.5}$ \citep{Kainulainen09a,Ballesteros-Paredes11a,Lombardi10a,Collins12a,federrath12,Kainulainen13b,Girichidis2014,MyersP2015,schneider2015MNRAS.453L..41S,Stutz2015A&A...577L...6S,Burkhart2015,Imara2016,padoan2017ApJ...840...48P,MyersP2017,Bialy2017ApJ...843...92B,Chen2017,Burkhart2017ApJ...834L...1B,2017ApJ...838...40M,2018arXiv180105428B}. These studies have noted that for gas with a density radial profile $\rho\propto r^{-\alpha}$, the corresponding PDF for $\log\rho$ is a powerlaw that scales as $\rho^{-3/\alpha}$. Thus, the collapse is consistent with the picture that dense shocks collapse into isothermal cores: $\rho^{-2}$.

The conceptual picture we have built up from the turbulent numerical simulations presented here supports the ansatz of
\cite{2014ApJ...785...69C,2015ApJ...810..126C}, which considered the formation of prestellar cores in a scenario that assumed a converging large-scale flow and shocked layer as an initial condition, with strongly magnetized regions that are locally sheetlike. Those studies found anisotropic contraction of cores along field lines, and also confirms our results that ambipolar diffusion is not necessary to form low-mass supercritical cores.

\section{Conclusions}
\label{sec:conc}

Astrophysical systems, such as molecular clouds out of which stars form, often show supersonic turbulent motions in a magnetized environment. The supersonic velocities of the fluid lead to shock formation, which are responsible for the densest regions in the fluid. These regions grow as they sweep up mass and can undergo gravitational collapse if they are not first dispersed on a sound crossing time  or magnetosonic crossing time (for perpendicular shocks).

We have identified and traced the dense regions in supersonic, isothermal turbulence that are sites of star formation.  These post-shock regions have densities enhanced by a factor of $\mathcal{M}^2$ (i.e., the isothermal jump conditions). Pre-stellar cores undergoing collapse are found to be associated with dense shocks that sweep up material and have an exponential density profile. In the case of strong magnetic fields, these shocks are strongly directed parallel to the magnetic field. Collapse along the direction parallel to the magnetic field is unhindered, but magnetic pressure and tension make collapse in the perpendicular direction difficult, requiring magnetic flux loss through an additional process, such as fast turbulent reconnection or ambipolar diffusion.

As sites of star formation in the supersonic turbulence model would be associated with strong shocks, it would mean young collapsing cores would be associated with complex, discontinuous magnetic field morphologies and filamentary strands (e.g. as seem in recent Atacama Large Millimeter/submillimeter Array (ALMA) observations of polarized dust emission from some Type 0 protostellar sources \citep{2017ApJ...842L...9H}.
The shocks, which sweep up mass and initially grow self-similarly until Jeans unstable, are part of the in situ star formation process and may have an effect on subsequent stellar evolution and resulting feedback.

\section*{Acknowledgments}
Support (PM) for this work was provided by NASA through Einstein Postdoctoral Fellowship grant number PF7-180164 awarded by the \textit{Chandra} X-ray Center, which is operated by the Smithsonian Astrophysical Observatory for NASA under contract NAS8-03060.
 BB acknowledges support from the Institute for Theory and Computation (ITC) Harvard-Smithsonian Center for Astrophysics Postdoctoral Fellowship.
Some of the computations in this paper were run on the Odyssey cluster supported by the FAS Division of Science, Research Computing Group at Harvard University.
PM would like to thank Jim Stone for valuable discussions on the subject of this manuscript.
\bibliography{mybib}{}

\bsp
\label{lastpage}
\end{document}